\documentclass{aa}  

\usepackage{graphicx}
\usepackage{color}
\usepackage{tabularx}
\usepackage{txfonts}
\usepackage[dvipsnames]{xcolor}
\usepackage[normalem]{ulem}
\usepackage[utf8]{inputenc}
\usepackage[T1]{fontenc}
\usepackage{mathtools}
\usepackage[thinc]{esdiff}
\usepackage{adjustbox}
\usepackage{float}
\usepackage{placeins}
\usepackage{stfloats}

\usepackage{nidanfloat}

\graphicspath{ {images/} }

\begin{document} 

   \title{Atomic diffusion in solar-like stars with MESA.}

   \subtitle{Comparison with the Montreal/Montpellier and CESTAM stellar evolution codes}

   \author{B. Campilho\inst{1,2}, M. Deal\inst{2} \and D. Bossini\inst{2} }

   \institute{Faculdade de Ciências da Universidade do Porto,
              Rua do Campo Alegre 1021 1055, 4169-007 Porto
              \and
              Instituto de Astrofísica e Ciências do Espaço, Universidade do Porto CAUP, Rua das Estrelas, PT4150-762 Porto, Portugal\\
              \email{up201704548@up.pt \\
              morgan.deal@astro.up.pt}}

   \date{Received (Date) / Accepted (Date)}

 
  \abstract
   {The stellar evolution code Modules for Experiments in Stellar Astrophysics (MESA) is public and is widely used by the community. It includes the possibility of taking several non-standard processes such as atomic diffusion into account. Even if the effect of gravitational settling is considered a standard ingredient in stellar modelling today, this is not the case for radiative accelerations. The specific treatment of atomic diffusion along with the radiative accelerations has never been compared with other stellar evolution codes. Benchmarking these codes is important because improved accuracy is required in order to analyse data from present and future space missions, such as the \textit{Kepler}, Transiting Exoplanet Survey Satellite (TESS), and PLAnetary Transits and Oscillations of stars (PLATO) missions.}
   {The aim of this paper is to compare MESA models including atomic diffusion (with radiative accelerations) with models computed with the Montreal/Montpellier stellar evolution code and with the Code d'Evolution Stellaire Adaptatif et Modulaire (CESTAM). Additionally, we assess the impact of some MESA options related to atomic diffusion.}
   {We calculated atomic diffusion, including radiative accelerations, following the abundance profiles of 14 elements with MESA models. This was then compared with 1.1 and 1.4~$M_{\odot}$ models computed with the Montreal/Montpellier and CESTAM codes. Various tests of MESA options for atomic diffusion were also carried out by varying only one of them at a time.}
   {We find that the abundance profiles of the considered elements in the MESA models compare rather well with the models computed with the two other codes when atomic diffusion options are carefully set. We also show that some options in MESA are crucial for a proper treatment of atomic diffusion.}
   {}


   \maketitle
%

\section{Introduction}

The transport of chemical elements inside stars plays a fundamental role in determining their structure and evolution. The distribution of chemical elements in stellar interiors is the result of complex interactions between macroscopic and microscopic processes, which are still not fully understood and therefore poorly modelled.

The proper modelling of the transport of chemical elements is becoming crucial for characterising stars, taking advantage of the high-quality asteroseismic data obtained by space missions such as \textit{Kepler} \citep{borucki10} or the Transiting Exoplanet Survey Satellite \citep[TESS,][]{ricker15} mission, as well as for future such missions.
As an example, the PLAnetary Transits and Oscillations of stars (PLATO) mission, which is an ESA space mission set to launch in 2026, aims to detect Earth-like planets in the habitable zone of Sun-like stars. 
In order to characterise the detected exoplanets, it is crucial to determine the properties of the host star (namely its mass, radius, and age) with high accuracy. The goal of this mission is to reach an accuracy of about or lower than 3\% in radius and 10\% in mass and age for the planets. This translates into the need of reaching accuracies of about or lower than 2\% in radius, 15\% in mass, and 10\% in age for host stars similar to the Sun \citep{rauer14}. However, this error budget can easily be exceeded when some transport processes (e.g. atomic diffusion) are neglected, especially for the age \citep[e.g.][]{nsamba18,deal_2018,aerts18,deal_rot}.

Atomic diffusion selectively affects chemical elements, and is a direct consequence of internal gradients in stars, such as pressure and temperature. Helioseismology has shown in the past decades that atomic diffusion must be included in solar models \citep{christensen93}. Other examples of cases in which atomic diffusion is important are in G-, F-, and A-type main-sequence stars, where not taking it into account has a significant effect on their structure and abundance profiles \citep[e.g.][]{theado09,richer00,richard01,michaud11,deal16,deal_2018}. When correctly modelled, atomic diffusion can be used to infer the efficiency of competing macroscopic transport processes \citep[][]{richard05,michaud11,verma19,deal_rot,semenova20}.

The actual diffusion velocity of an element mainly depends on the competition between two main forces (or accelerations): gravity, and radiative accelerations ($g_\mathrm{rad}$). The latter is a transfer of momentum between photons and ions and counteracts gravitational settling. It is different for each element. Thus, diffusion has a direct effect on the abundance profiles in the star, which in turn leads to modifications in the Rosseland opacities. 

Atomic diffusion is efficient in the outer layers of stars because the diffusion timescale is approximately proportional to the density of protons \citep{deal_2018}. Therefore, accurate predictions of surface abundances, which are necessary for the inference of stellar properties, also require taking into account atomic diffusion, including radiative accelerations \citep[e.g.][]{richard01,michaud11}.  Surface metallicity is another fundamental stellar parameter that is affected by atomic diffusion, and it is important for the study of exoplanetary systems because the probability of a star to have a detectable planetary system is linked to its metal content \citep[e.g.][]{alecian}. 

The number of stars for which atomic diffusion (including radiative accelerations) has an important effect ranges from 33\% up to 59\% of the PLATO core program star sample \citep{deal_2018}. This means that for a large number of the stars observed by PLATO, an accurate modelling of atomic diffusion will be necessary to achieve the requirements of the mission.


Despite their importance, only a few evolution codes currently incorporate consistent computations of stellar models that include the complete treatment of atomic diffusion. The code Modules for Experiments
in Stellar Astrophysics \citep[hereafter MESA;][]{paxton,paxton2} is one of them. The aim of this paper is to compare \textsc{MESA} models with models computed using the Montreal/Montpellier evolution code \citep{turcotte98,richer00} and the Code d'Evolution Stellaire Adaptatif et Modulaire  \citep[hereafter CESTAM, the "T" stands for transport;][]{morel08,marques13,deal_2018}, all models including atomic diffusion with the contribution of the radiative accelerations.

The first objective is to show how \textsc{MESA} compares to the two other codes at a given input physics (these two other codes were already compared, giving very similar results, in \citealt{deal_2018}). The second objective is to provide a comprehensive assessment of the implications of using the different MESA atomic diffusion options dedicated to speed up these computations with some approximations, and pointing out those that may affect the resulting models most.

This paper is organised as follows: in Sects. 2 and 3, we present a brief introduction to atomic diffusion and the way in which \textsc{MESA}, \textsc{CESTAM,} and the Montreal/Montpellier code calculate it, as well as a direct comparison between them. Then, in Sect. 4, we present the comparison between the three codes. Section 5 addresses the impact of \textsc{MESA} atomic diffusion options that may have a significant effect on the resulting abundance profiles, which is then continued in Appendix~\ref{Appendix_Opt}. Finally, Sect. \ref{conc} is dedicated to the conclusion. 

\section{Atomic diffusion}

Atomic diffusion is calculated from first principles of physics, being a direct consequence of the internal gradients of pressure, temperature, and concentration in stars. This can be seen in the diffusion equation,
\begin{equation}
\label{diff_eq}
     \rho \frac{\partial X_{i}}{\partial t} = \frac{1}{r^2} \frac{\partial }{\partial r} \left[r^2 \rho D_\mathrm{turb} \frac{\partial X_{i}}{\partial r} \right] - \frac{1}{r^2} \frac{\partial }{\partial r} \left[r^2 \rho v_i \right] + \\
     A_i m_p \left[\sum_{j}^{} (r_{ji} - r_{ij})\right]
 ,\end{equation}
 
\noindent where $X_i$ is the mass fraction of element $i$, $v_i$ is its atomic diffusion velocity, $\rho$ is the density in the considered layer, $D_\mathrm{turb}$ is the turbulent diffusion coefficient (i.e. the combined effect of the macroscopic transport processes, which cannot be simply treated as their sum due to their possible couplings and interactions), $A_i$ is its atomic mass, $m_p$ is the mass of a proton, and $r_{ij}$ is the reaction rate of the reaction that transforms element $i$ into $j$.
 
Additionally, the diffusion velocity of a trace element $i$ is given by
 
\begin{eqnarray}
\label{dif_v_eq}
    v_i = D_{ip} \left[-\frac{\partial X_i}{\partial r} + \frac{A_i m_p}{kT}(g_{\mathrm{rad},i} - g) + \frac{(\overline{Z_i} + 1)m_pg}{2kT} + \kappa_T \frac{\partial ln T}{\partial r} \right]
 ,\end{eqnarray}

\noindent where $D_{ip}$ is the diffusion coefficient of element $i$ in a proton dominated plasma. The variable $g$ is the local gravity,
$\overline{Z_i}$ is the average charge (in proton charge units) of the element $i$ (roughly equal to the charge of the dominant ion), $k$ is the Boltzmann constant, $T$ isthe temperature, $\kappa_T$ is the thermal diffusivity, and $g_{\mathrm{rad},i}$ is the radiative acceleration on element $i$  \citep{richer98},

\begin{equation}
\label{g_rad_eq}
     g_{\mathrm{rad},i} = \frac{1}{4 \pi r^2} \frac{L_{r}^\mathrm{rad}}{c} \frac{\kappa_R}{X_i} \int_{0}^\infty \frac{\kappa_{u, i}}{\kappa_{u,\mathrm{total}}} \mathcal{P}(u) {\rm d}u \hspace{2pt,}
 \end{equation}
 
\noindent where $L_{r}^{rad} /(4 \pi r^2 c)$ is the total radiative momentum flux at radius r, u $\equiv h\nu/(kT)$ is the dimensionless frequency variable, $\mathcal{P}(u)$ is the normalised blackbody flux, $\kappa_{u,i}$ is the monochromatic opacity of element $i$, $\kappa_{u,\mathrm{total}}$ is the total monochromatic opacity, and $\kappa_R$ is the Rosseland opacity, which can be obtained by integrating over the spectrum,
 \begin{equation}
 \label{ross_op_eq}
     \frac{1}{\kappa_{R}} = \int{\frac{1}{\kappa_{\nu}} \frac{{\rm d}B_{\nu}}{{\rm d}T} {\rm d}\nu} \;  \bigg/ \int{\frac{{\rm d}B_\nu}{{\rm d}T}{\rm d}\nu}.
 \end{equation}

Therefore, the diffusion velocity of an element depends mainly on two opposite forces (or accelerations): gravity, which causes the element to migrate towards the centre of the star (gravitational settling), and radiative accelerations. Radiative accelerations result from a selective absorption by the various chemical species present in the star of part of the net outgoing momentum flux carried by photons. Usually, this absorption produces a net outward force on each absorbing species, which can exceed its weight. If radiative accelerations are present, both over- and underabundances (compared to the initial abundance) appear in the abundance profiles of the star, mostly as a result of the imbalance between radiative forces and gravity. The importance of radiative accelerations comes from the fact that they dominate atomic diffusion in many stars, especially in the hottest white dwarfs and in main-sequence stars with $T_{eff} > 6000 K$ \citep{richer98}. 
Nevertheless, competing hydrodynamic processes \cite[e.g. rotationally induced mixing;][]{deal_rot} may moderate the expected effects of atomic diffusion.




\section{Stellar evolution codes}
\label{sec:codes}

There are different sets of equations to solve the diffusion equations and different approaches for calculating $g_\mathrm{rad}$. Some give priority to either accuracy or computation time, while others try to find a compromise between them \citep[for an overview, see][]{alecian18}. In this section we present the methods used by three codes, \textsc{MESA}, Montreal/Montpellier and \textsc{CESTAM,}  to treat atomic diffusion, including $g_\mathrm{rad}$, before summarising the differences between them.

\subsection{MESA}
\label{MESA}
 
Atomic diffusion is calculated in \textsc{MESA} by solving the Burgers equations \citep{burgers} using the \cite{thoul} formalism, while radiative accelerations are calculated using the OPCDv3.3 package \citep{seaton052}. It follows an opacity sampling method with opacity tables at fixed frequency grids. In order to compute the integral of Eq.~\ref{g_rad_eq}, \textsc{MESA} includes the modifications of the original OPCD routines described in \cite{hu11}. Radiative accelerations are computed for H, He, C, N, O, Ne, Na, Mg, Al, Si, S, Ca, Fe, and Ni (see Sect. \ref{ni} for the case of nickel). 
This method is less accurate than the direct use of atomic data, but it has the advantage of being faster to compute.
\textsc{MESA} also offers a vast number of customizable options, including options for atomic diffusion. In Appendix \ref{Appendix_Opt} we focus on some of them. A more detailed description of the treatment of atomic diffusion in \textsc{MESA} can be found in \cite{paxton1,paxton}.


\subsection{Montreal/Montpellier code} 
 
In the  Montreal/Montpellier, atomic diffusion is calculated by solving the full set of Burgers  equations \citep{burgers} for H, He, C, N, O, Ne, Na, Mg, Al, Si, P, S, Cl, Ar, K, Ca, Ti, Cr, Mn, Fe, and Ni.
Radiative accelerations are also computed with an opacity sampling method using the OPAL monochromatic opacity table \citep[not publicly available]{iglesias}, with a fixed frequency grid. 
A more detailed description of the treatment of atomic diffusion in the Montreal/Montpellier code can be found in \cite{turcotte98} and \cite{richer98}.
 
\subsection{CESTAM}
 
In the \textsc{CESTAM} code, atomic diffusion is computed following the formalism presented in \cite{Michaud_f2}, which is an approximation of the Burgers equations, for H, He, C, N, O, Ne, Na, Mg, Al, Si, S, Ca, and Fe. Atomic diffusion can also be computed solving the full set of Burgers equations \citep{burgers}. The models used in this study are computed with the first option. Radiative accelerations are computed using the single-valued parameter (SVP) derived in \cite{leblanc_alecian}, which is based on a simplified form of Eq.~\ref{g_rad_eq} obtained by separating the terms involving the atomic quantities from those describing the local plasma. The use of the resulting parametric equations allows for faster computations when compared to opacity sampling because it involves very short pre-calculated tables. 
The SVP method is only valid for stellar interiors. It has to be noted that the \textsc{CESTAM} models presented in this study take the latest version of the SVP method into account \citep{alecian20}. A more detailed description of the treatment of atomic diffusion in \textsc{CESTAM} can be found in \cite{morel08} and \cite{deal_2018}.

\subsection{Differences between the three codes}
\label{diff_codes}
The treatment of atomic diffusion is similar in the three codes, being based on the Burgers equations, with slight differences: \textsc{MESA} uses the \cite{thoul} formalism and CESTAM the \cite{Michaud_f2} formalism. It has already been shown by code comparisons within the CoRoT/ESTA program that these formalisms give very similar results \citep{lebreton08}. To compute radiative accelerations, \textsc{MESA} and the Montreal/Montpellier code both use the opacity sampling method, while CESTAM uses the SVP approximation. The SVP method is numerically faster, but introduces a maximum uncertainty of approximately 30\% on the $g_\mathrm{rad}$ \citep{alecian20}. We show in the next sections that this uncertainty is generally lower.

In addition to the different computation methods, the codes also use different opacity data. \textsc{MESA} and the SVP approximation implemented in CESTAM use the Opacity Project (OP) data, while the Montreal/Montpellier code uses the OPAL data. Nevertheless, the difference between the OPAL and OP data is close in terms of Rosseland mean opacities \citep[e.g. 5\% at the bottom of the convective zone of the Sun,][]{seaton04}. Therefore, we expect the differences in the radiative acceleration profiles to rather come from the difference in the monochromatic opacity of each element (integral of Eq.~\ref{g_rad_eq}) than from the differences in the Rosseland mean opacities between OP and OPAL. We also expect the abundance profiles of the \textsc{MESA} models to be closer to those of CESTAM because radiative accelerations are computed from the same opacity data (OP).

\section{Comparison of MESA models with other codes} \label{comp_section}

In this section, we first present the physical input parameters of the models computed with the three evolution codes (\textsc{MESA}, Montreal/Montpellier, and \textsc{CESTAM}). Then, we compare their internal abundance profiles and surface abundances, before proceeding to analyse the particular case of nickel in more detail.

\subsection{Model input physics} 

\begin{figure*}
    \centering
      \includegraphics[scale=0.7]{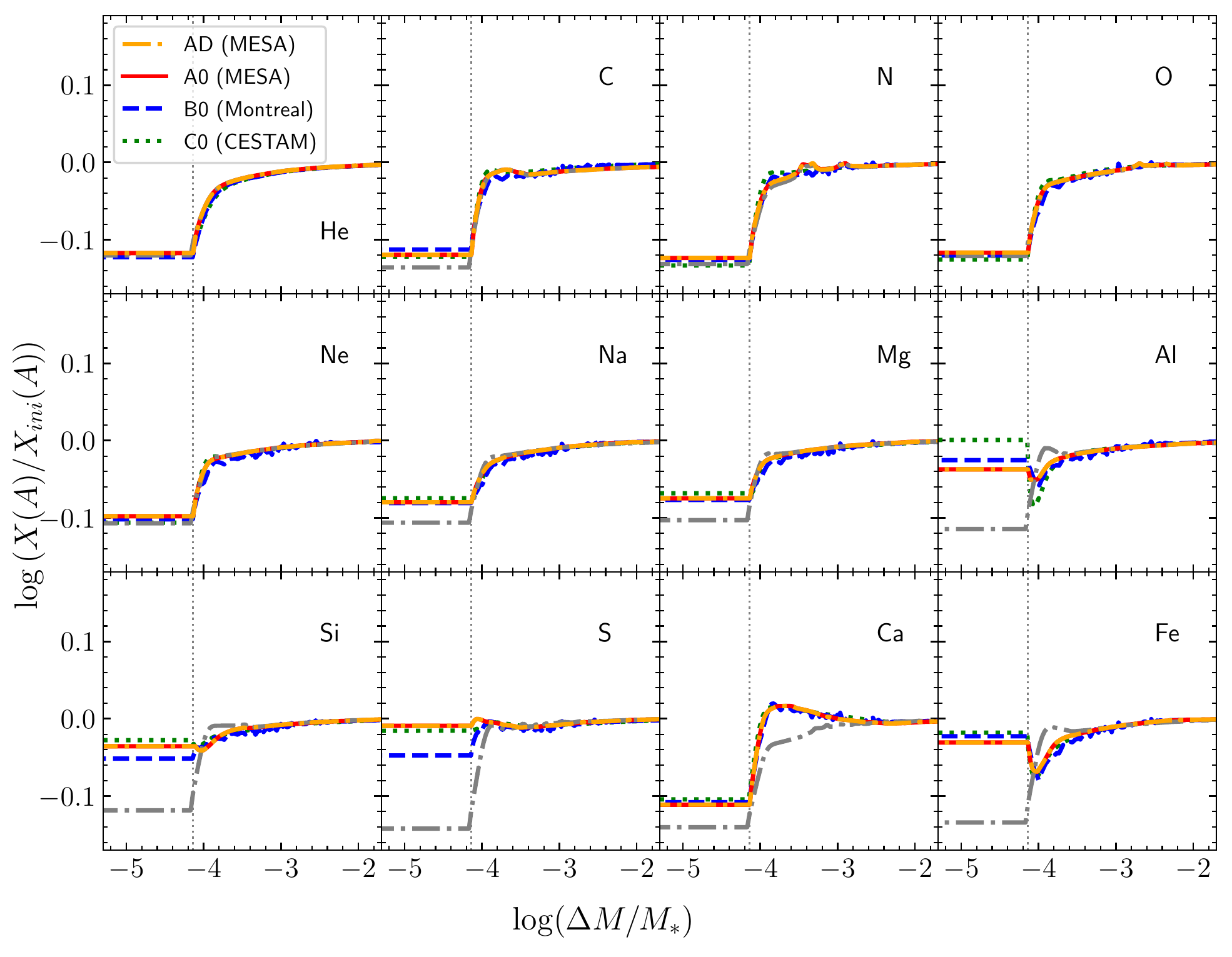}
      \caption{Abundance profiles according to $\log(\Delta M/M_\ast)$ (with $\Delta M$ being the mass above the radius $r$) obtained with \textsc{MESA} (Ad model, dot-dashed orange lines; A0 model, solid red lines), the Montreal/Montpellier code (B0 model, dashed blue lines) and \textsc{CESTAM} (C0 model, dotted green lines) at 420~Myr. The dot-dashed grey lines represent the abundance profiles for the Ad model without radiative accelerations. The vertical dotted grey line represents the average position of the surface convective zone of the three models. The surface is towards the left part of the plots.}
   \label{compar}
\end{figure*}


In order to perform our comparison, we selected models of two different masses (1.1 and 1.4~$M_\odot$).
The mass of 1.1~$M_\odot$ was chosen because it develops a deeper surface convective zone (SCZ) during the main sequence, mainly to compare the effect of gravitational settling, while the 1.4~$M_\odot$ model allows us to make a direct comparison with the models in \cite{deal_2018}. 
The surface convective zone of these models is small enough to make the effects of radiative accelerations noticeable, but not too small to prevent unrealistic abundance variations appearing in models that neglect known additional transport processes, such as rotation-induced mixing.

All the models used for this comparison, as well as their selected physical input parameters, are presented in Table \ref{tab1}:\\

\textbf{Montreal/Montpellier models}

The reference models for the comparisons are computed with the Montreal/Montpellier evolution code. The 1.4~M$_\odot$ model (B0) is the same as was used in the comparison with CESTAM model in \cite{deal_2018}.\\

\textbf{MESA models}

\textsc{MESA} models are computed with the same initial chemical composition, metal mixture \citep{Grevesse_1993}, and atmospheres (Eddington) as the Montreal/Montpellier models. The main differences are in the monochromatic opacity tables, the equation of states, and the nuclear reaction rates. We also calibrated the value of $\alpha_\mathrm{MLT}$ in order to obtain the position of the bottom of the surface convective zone as close as possible to the Montreal/Montpellier models. This step is necessary in order to compare the abundance profiles and distinguish the differences coming from the structure and coming from the opacity data used to compute radiative accelerations. The monochromatic opacities were considered for a temperature range [$10^4$,$10^{6.3}$]~K, using options 3, 4, 5, and 6 of Table~\ref{tab_tests}. The default model (AD) includes the default value for atomic diffusion options, which speeds up the computations, while the other models (A0 and A1) include an optimised setup for some of these options. These specific options are discussed in Sect. \ref{MESA_opsect} and are detailed in Appendix~\ref{Appendix_Opt}. The \textsc{MESA} model A0\footnote{\url{http://cococubed.asu.edu/mesa_market/inlists.html}} (see Table~\ref{tab1}) was used as a reference to test the impact of the different atomic diffusion options available in \textsc{MESA}.\\

\textbf{CESTAM models}

The CESTAM models are computed with a more recent version of the code than was used in the \cite{deal_2018} comparison. This version of the code includes the improved method described in \cite{huibonhoa21} to calculate the Rosseland mean opacity with the OPCD monochromatic opacity tables \citep{seaton052}. It also includes the latest version of the SVP tables \citep{alecian20}. Similarly to the \textsc{MESA} models, $\alpha_\mathrm{MLT}$ was adjusted, and the same temperature range was defined for the use of monochromatic opacities to compute the Rosseland mean opacity.



\begin{table}[]
\caption{Input parameters of the compared models.}
\label{tab1}
\begin{adjustbox}{width=9cm,center}
\begin{tabular}{l|cc|cc|cc}
\hline
\hline
Code          & \multicolumn{2}{c}{\textsc{MESA}} & \multicolumn{2}{c}{MoMo}  & \multicolumn{2}{c}{CESTAM}   \\
\hline 
Model          &  AD/A0 &  A01  & B0 & B01  &  C0 &  C01   \\
\hline 
Mass [$M_{\odot}$] & 1.4 & 1.1  & 1.4 & 1.1 & 1.4 & 1.1    \\
\hline 
Age (Myrs) & 420 & 3050  & 420 & 3050 & 420 & 3050 \\
\hline
$X_{ini}$ & \multicolumn{6}{c}{0.69500}      \\
$Y_{ini}$ & \multicolumn{6}{c}{0.27995}       \\
$(Z/X
)_{ini}$ & \multicolumn{6}{c}{0.0360}  \\
Core ovs. & \multicolumn{6}{c}{None}             \\
Atmosphere & \multicolumn{6}{c}{Eddington}    \\
Mixture & \multicolumn{6}{c}{GN93 \tablefoottext{a}}\\
\hline
$\alpha_{MLT}$ & 1.717 & 1.721 & \multicolumn{2}{c|}{1.687} &  1.682 & 1.700 \\
\hline
Opacities  & \multicolumn{2}{c|}{OPCD+OPAL} & \multicolumn{2}{c|}{OPAL Mono}  & \multicolumn{2}{c}{OPCD+OPAL}  \\
EoS  & \multicolumn{2}{c|}{OPAL2005\tablefoottext{b}} & \multicolumn{2}{c|}{CEFF\tablefoottext{c}} & \multicolumn{2}{c}{OPAL2005} \\
Nuc. React. & \multicolumn{2}{c|}{NACRE\tablefoottext{d}} & \multicolumn{2}{c|}{Bahcall92\tablefoottext{e}}   & \multicolumn{2}{c}{NACRE} \\
\hline
\end{tabular}
\end{adjustbox}
\tablefoot{\tablefoottext{a}{\cite{Grevesse_1993}},\tablefoottext{b}{\cite{rogers02}},\tablefoottext{c}{\cite{JCD92}}, \tablefoottext{d}{\cite{angulo99}},\tablefoottext{e}{\cite{bahcall92}}}
\end{table}

\subsection{Comparison of abundance profiles}
\label{ab_compar}

The abundance profiles of $^4$He, $^{12}$C, $^{14}$N, $^{16}$O, $^{20}$Ne, $^{23}$Na, $^{24}$Mg, $^{27}$Al, $^{28}$Si, $^{32}$S, $^{40}$Ca, and $^{56}$Fe are presented in Fig. \ref{compar} for the 1.4$M_{\odot}$ models. As expected, the agreement between A0 and B0/C0 is very satisfactory. The depletion in $^4$He, $^{12}$C, $^{14}$N, $^{16}$O, $^{20}$Ne, $^{23}$Na, $^{24}$Mg, and $^{40}$Ca, as well as the accumulations of $^{27}$Al and $^{56}$Fe, are well reproduced by the A0 model. However, the profiles of $^{28}$Si and $^{32}$S in the A0 model are not consistent with the B0 model, while they are consistent with the C0 model (with larger differences in the case of $^{27}$Al). As we explain below, this is mainly due to the differences in the monochromatic opacity data used by the codes (Table~\ref{tab1}). The AD and A0 models show identical abundance profiles. This shows that when they are properly set up, the options allowing us to speed up atomic diffusion computations have no effect on the abundance profiles (see Appendix~\ref{Appendix_Opt} for more details about these options).

\begin{figure}
    \centering
     \includegraphics[scale=0.6]{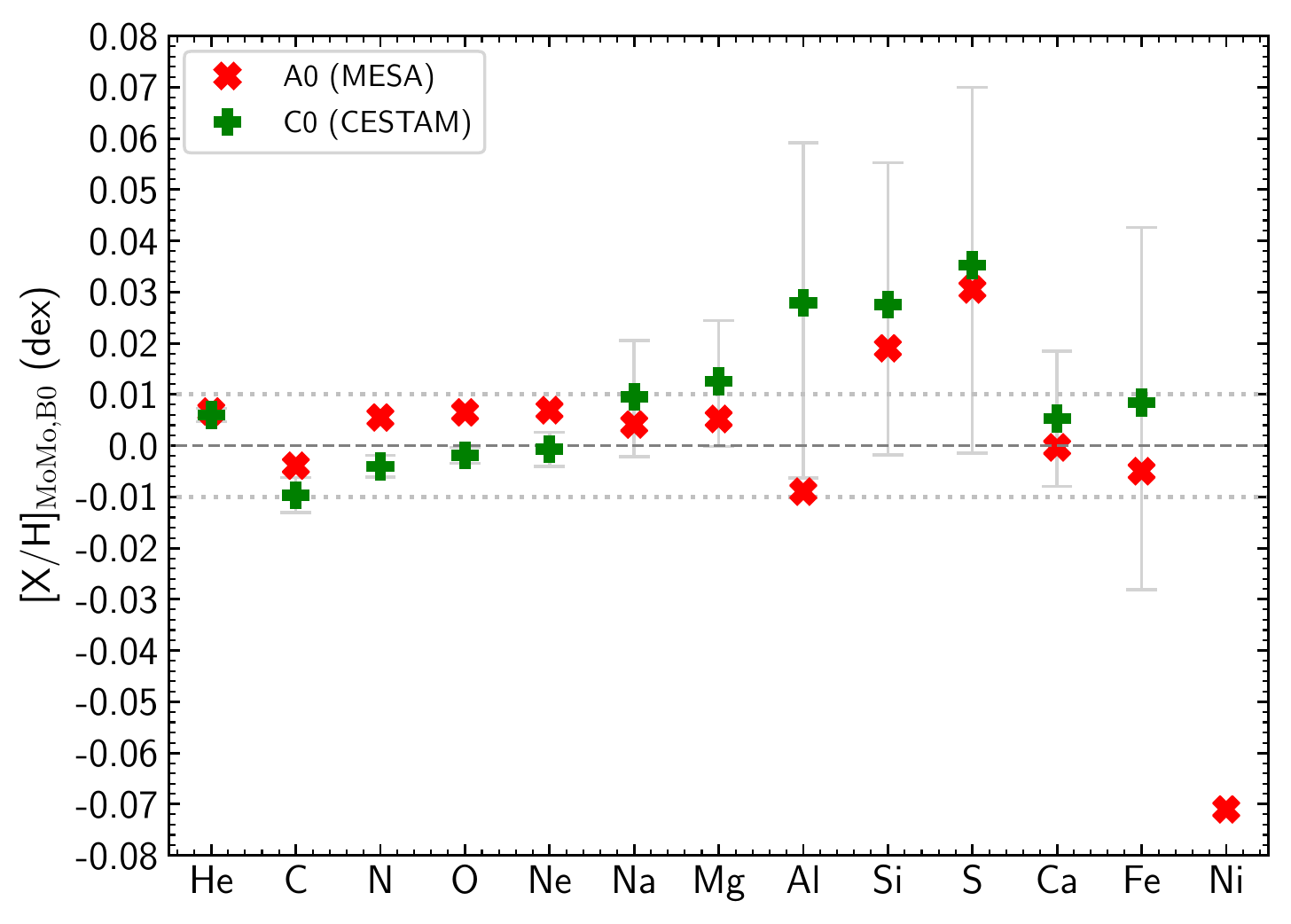}
      \caption{Difference of surface abundances$^2$ between the model B0 and the 2 other models (A0: red symbols, and C0: green symbols) shown in Fig.~\ref{compar}. The error bars of the C0 model are estimated using the 30\% uncertainties on the radiative accelerations values obtained with the SVP method. The dotted grey lines show the $\pm0.01$~dex interval.}
   \label{1.4-reldiff}
\end{figure}

When comparing all the curves with the dot-dashed grey lines (model without radiative accelerations), we clearly see that the effect of radiative accelerations is different from one element to the next. For He, N, O, and Ne, in this specific case, radiative accelerations are negligible.

Figure~\ref{1.4-reldiff} shows the difference in surface abundances between the Montreal/Montpellier model (B0) and the \textsc{MESA} (A0) and CESTAM (C0) models\footnote{[A/H]$_\mathrm{MoMo}=\log_{10}\left(\frac{X(A)}{X(H)}\right)_\mathrm{MESA/CESTAM}-\log_{10}\left(\frac{X(A)}{X(H)}\right)_\mathrm{MoMo}$, with $X(A)$ and X(H) the mass fractions of the element A and hydrogen, respectively.}. The differences in temperature, density, gravity, and Rosseland mean opacity around the bottom of the surface convective zones are smaller than 3\%, 4\%, 1\% and 1\%, respectively. The differences in the gravity profiles (in direct competition with radiative accelerations) cannot induce large differences in the abundance profiles in these conditions (see Fig.~\ref{compar-grad}, for which the gravity profile of the three models overlap). It leads to surface abundance differences lower than 0.01~dex for the elements not affected by radiative accelerations (e.g. He, C, and O). For almost all of the elements in the \textsc{MESA} model, the differences are smaller, except for $^{28}$Si and $^{32}$S. 

For $^{28}$Si and $^{32}$S, this is the indication that these differences come from the monochromatic opacity data, affecting the integral in Eq.~\ref{g_rad_eq}. The radiative acceleration profiles for the A0 and C0 models are close but slightly larger than gravity at the bottom of the SCZ, while this is the opposite for the B0 model (see Fig.~\ref{compar-grad}). This explains why these two elements are accumulated in the A0 and C0 models, while that is not the case with the B0 model. The comparison with the CESTAM model (C0) confirms the origin of the behaviour of $^{28}$Si and $^{32}$S, as both codes use the same opacity data to compute radiative accelerations and show the same behaviour. In the case of $^{32}$S, this difference can be seen in Fig. 7 of \cite{seaton04}, where its contribution to the Rosseland mean opacity at the temperature of the BSCZ is slightly higher with OP than OPAL opacities, which leads to slightly higher values of the radiative acceleration for this element.

For $^{27}$Al, the surface abundance of the C0 model is different from that of A0 (by $\approx0.03~dex$) despite the fact A0 and B0 are close. In this case, the difference cannot come from the monochromatic opacity data. On the other hand, the radiative acceleration profiles shown in Fig.~\ref{compar-grad} indicate that the C0 model is closer to the B0 models than to that of A0. Nevertheless, the $^{27}$Al abundance of the C0 model is consistent with the A0 and B0 abundances when considering the 30\% uncertainty of the SVP method. This behaviour of the $^{27}$Al abundance for A0 and C0 may be a combination of the effect of evolution and the radiative acceleration computation. Moreover, most of the time, surface abundances predicted with the SVP approximation are very close to those of \textsc{MESA.}  This indicates that the 30\% uncertainties of this method (see Sect.~\ref{diff_codes}) are often smaller, and it confirms the robustness of the method. 

The abundance profiles are even closer for the 1.1~$M_{\odot}$ models (see Fig. \ref{compar-1.1}). In this case, gravitational settling dominates the transport (which is less subject to uncertainties than radiative accelerations), hence the smaller differences. Their average differences are smaller than 0.005~dex (Fig.~\ref{compar-1.1v2}). This also confirms the origin of the behaviour of $^{28}$Si and $^{32}$S in the A0 and C0 models compared to that of B0.

\subsection{Case of nickel}\label{ni}

\begin{figure}[ht]
    \centering
    \includegraphics[scale=0.57]{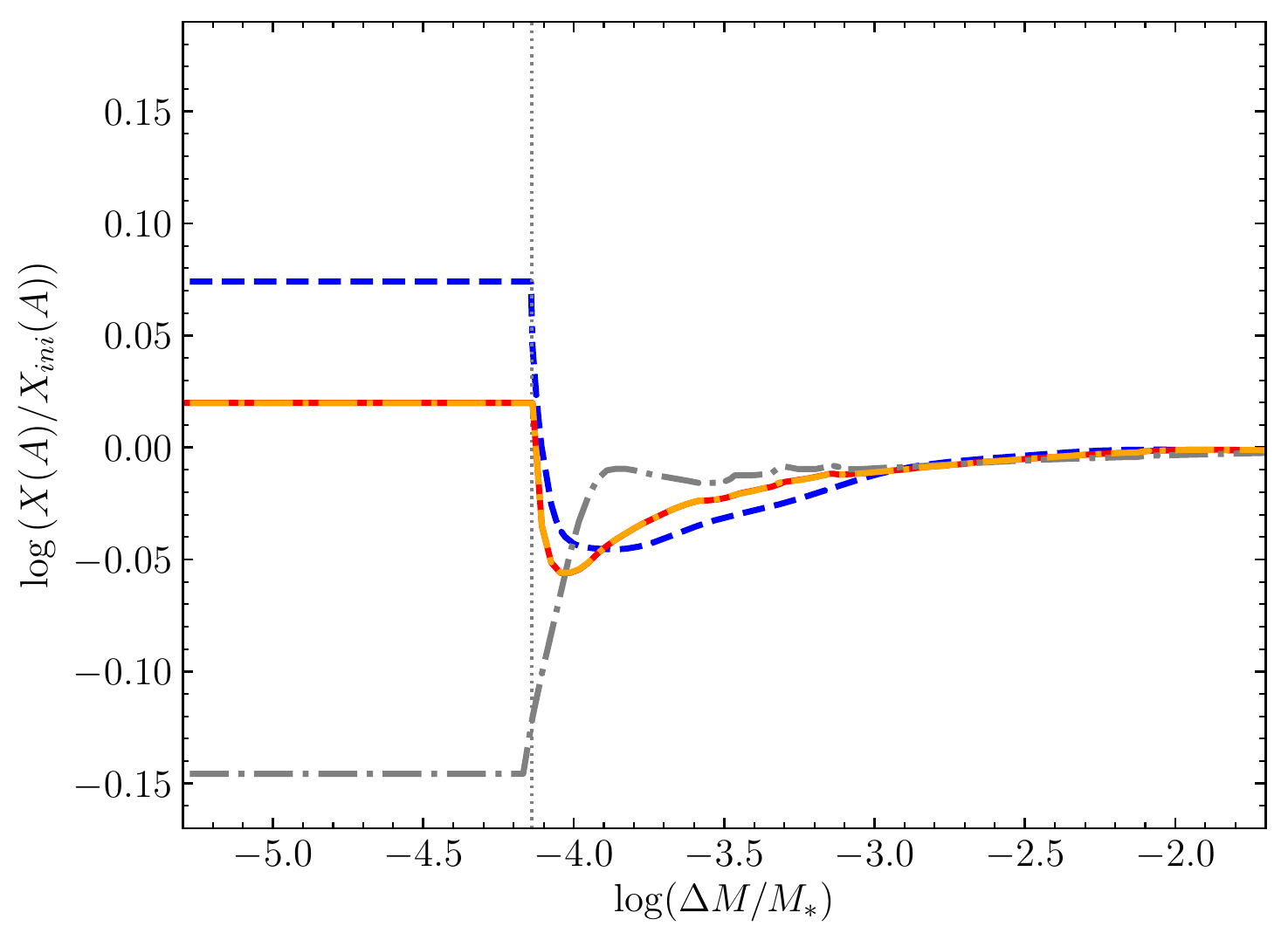}
    \caption{Same legend as Fig. \ref{compar} for nickel.}
   \label{fig:Ni}
\end{figure}

The nickel monochromatic opacity data available in the OPCD package are not determined similarly to the other elements, but are derived from those of iron. Figure \ref{fig:Ni} shows the comparison of the abundance profile of nickel between the A0 and B0 models. Even though the abundance profile pattern is similar, the surface abundances of the two models are significantly different (10\% of the surface mass fraction). 
Similarly to the previous section, Figure~\ref{1.4-reldiff} shows that the differences in the predicted Ni surface abundance ($\approx 0.07$~dex) cannot be related to the structure differences, are more than two times larger than the effect of the difference in the opacity data we see for $^{28}$Si and $^{32}$S, and are larger than the difference of $^{27}$Al in the C0 model.
For this reason, the surface abundances of nickel predicted by \textsc{MESA} models, and more generally by the OPCD package, should be taken with caution.
This is also the reason why nickel is currently not available in C0 and \textsc{CESTAM} models (treated as an element at equilibrium) using the SVP method \citep{alecian20}.

\section{Atomic diffusion options in MESA}
\label{MESA_opsect}

In this section we briefly address the impact of atomic diffusion options dedicated to computational optimisation available in \textsc{MESA}, which may have a significant effect on the resulting abundance profiles unless they are carefully set up. Quantitative results, along with a more detailed explanation of the options and their effects on surface abundances, are presented in Appendix \ref{Appendix_Opt}.

Two of the most important options allow consideration of a region at the surface as a single cell for atomic diffusion calculations (options 1 and 2 of Table~\ref{tab_tests}). The size of this region is defined either by a mass fraction of the total mass or by a specific temperature. While this helps minimise the computation time, these values must be chosen thoughtfully. If this value is higher than the fraction of mass of the SCZ, errors may occur. In this case, the mixed region is artificially forced to be deeper, introducing an artificial and efficient full mixing in regions where it should not be present. This option is further analysed in Appendix \ref{Appendix_Opt}, as well as other options whose effect on the surface abundances can be significant.


\section{Conclusion}
\label{conc}

We computed stellar models with the \textsc{MESA} evolution code taking atomic diffusion into account (including radiative accelerations). We compared these models to others computed with the Montreal/Montpellier code and \textsc{CESTAM}.

The models computed with the three codes compared in this study (\textsc{MESA}, Montreal/Montpellier, and CESTAM) show a very high level of agreement in terms of abundance profiles and surface abundances (Sect. \ref{ab_compar}). However, due to differences in the opacity data used by the Montreal/Montpellier code, its abundance profiles for $^{28}$Si and $^{32}$S show slight differences from those of \textsc{MESA} and CESTAM.

We showed that some options dedicated to speed up atomic diffusion computation in \textsc{MESA} need to be set carefully. The most important are options 1 and 2 of Table \ref{options}. These two options act on the treatment of atomic diffusion in the surface layers, making it possible to treat a certain region at the surface as a single cell for atomic diffusion calculations. This is very useful to speed up the computation, but leads to wrong abundance profiles when this region is larger than the natural surface convective zone. It mimics the addition of very efficient mixing at the bottom of the SCZ, modifying the predicted surface abundances. 
This also causes structural modifications, as the new chemical mixture changes the opacity in this zone, altering its size and the effective temperature.

The code comparison carried out in this paper showed that the \textsc{MESA} stellar evolution code compares well with the Montreal/Montpellier code and \textsc{CESTAM}. This kind of comparison and benchmarking of the codes is fundamental to assess the robustness of stellar evolution computation using different numerical methods, and atomic data for opacities and atomic diffusion calculation. It is also important in order to analyse present and future observations that missions such as \textit{Kepler}, TESS, or PLATO provide or will provide.

\begin{acknowledgements}
This work was supported by FCT/MCTES through the research grants UIDB/04434/2020, UIDP/04434/2020 and PTDC/FIS-AST/30389/2017, and by FEDER - Fundo Europeu de Desenvolvimento Regional through COMPETE2020 - Programa Operacional Competitividade e Internacionalização (grant: POCI-01-0145-FEDER-030389). MD and DB are supported by national funds through FCT in the form of a work contract. We thank an anonymous referee for valuable comments which helped to improve the paper.
 
\end{acknowledgements}  
\bibliographystyle{aa} 
\bibliography{ref.bib} 

\begin{thebibliography}{42}
\expandafter\ifx\csname natexlab\endcsname\relax\def\natexlab#1{#1}\fi

\bibitem[{{Aerts} {et~al.}(2018){Aerts}, {Molenberghs}, {Michielsen},
  {Pedersen}, {Bj{\"o}rklund}, {Johnston}, {Mombarg}, {Bowman}, {Buysschaert},
  {P{\'a}pics}, {Sekaran}, {Sundqvist}, {Tkachenko}, {Truyaert}, {Van Reeth},
  \& {Vermeyen}}]{aerts18}
{Aerts}, C., {Molenberghs}, G., {Michielsen}, M., {et~al.} 2018, \apjs, 237, 15

\bibitem[{{Alecian}(2018)}]{alecian18}
{Alecian}, G. 2018, in Astronomical Society of the Pacific Conference Series,
  Vol. 515, Workshop on Astrophysical Opacities, 169

\bibitem[{{Alecian} \& {LeBlanc}(2020)}]{alecian20}
{Alecian}, G. \& {LeBlanc}, F. 2020, \mnras, 498, 3420

\bibitem[{{Alecian, G.} \& {Michaud, G.}(2005)}]{alecian}
{Alecian, G.} \& {Michaud, G.} 2005, A\&A, 431, 1

\bibitem[{{Angulo}(1999)}]{angulo99}
{Angulo}, C. 1999, in American Institute of Physics Conference Series, Vol.
  495, American Institute of Physics Conference Series, 365--366

\bibitem[{{Bahcall} \& {Pinsonneault}(1992)}]{bahcall92}
{Bahcall}, J.~N. \& {Pinsonneault}, M.~H. 1992, Reviews of Modern Physics, 64,
  885

\bibitem[{{Borucki} {et~al.}(2010){Borucki}, {Koch}, {Basri}, {Batalha},
  {Brown}, {Caldwell}, {Caldwell}, {Christensen-Dalsgaard}, {Cochran},
  {DeVore}, {Dunham}, {Dupree}, {Gautier}, {Geary}, {Gilliland}, {Gould},
  {Howell}, {Jenkins}, {Kondo}, {Latham}, {Marcy}, {Meibom}, {Kjeldsen},
  {Lissauer}, {Monet}, {Morrison}, {Sasselov}, {Tarter}, {Boss}, {Brownlee},
  {Owen}, {Buzasi}, {Charbonneau}, {Doyle}, {Fortney}, {Ford}, {Holman},
  {Seager}, {Steffen}, {Welsh}, {Rowe}, {Anderson}, {Buchhave}, {Ciardi},
  {Walkowicz}, {Sherry}, {Horch}, {Isaacson}, {Everett}, {Fischer}, {Torres},
  {Johnson}, {Endl}, {MacQueen}, {Bryson}, {Dotson}, {Haas}, {Kolodziejczak},
  {Van Cleve}, {Chandrasekaran}, {Twicken}, {Quintana}, {Clarke}, {Allen},
  {Li}, {Wu}, {Tenenbaum}, {Verner}, {Bruhweiler}, {Barnes}, \&
  {Prsa}}]{borucki10}
{Borucki}, W.~J., {Koch}, D., {Basri}, G., {et~al.} 2010, Science, 327, 977

\bibitem[{{Burgers}(1969)}]{burgers}
{Burgers}, J.~M. 1969, {Flow Equations for Composite Gases} (Academic Press)

\bibitem[{{Christensen-Dalsgaard} \& {Daeppen}(1992)}]{JCD92}
{Christensen-Dalsgaard}, J. \& {Daeppen}, W. 1992, \aapr, 4, 267

\bibitem[{{Christensen-Dalsgaard} {et~al.}(1993){Christensen-Dalsgaard},
  {Proffitt}, \& {Thompson}}]{christensen93}
{Christensen-Dalsgaard}, J., {Proffitt}, C.~R., \& {Thompson}, M.~J. 1993,
  \apjl, 403, L75

\bibitem[{{Deal} {et~al.}(2018){Deal}, {Alecian}, {Lebreton}, {Goupil},
  {Marques}, {LeBlanc}, {Morel}, \& {Pichon}}]{deal_2018}
{Deal}, M., {Alecian}, G., {Lebreton}, Y., {et~al.} 2018, \aap, 618, A10

\bibitem[{{Deal} {et~al.}(2020){Deal}, {Goupil}, {Marques}, {Reese}, \&
  {Lebreton}}]{deal_rot}
{Deal}, M., {Goupil}, M.~J., {Marques}, J.~P., {Reese}, D.~R., \& {Lebreton},
  Y. 2020, \aap, 633, A23

\bibitem[{{Deal} {et~al.}(2016){Deal}, {Richard}, \& {Vauclair}}]{deal16}
{Deal}, M., {Richard}, O., \& {Vauclair}, S. 2016, \aap, 589, A140

\bibitem[{Grevesse \& Noels(1993)}]{Grevesse_1993}
Grevesse, N. \& Noels, A. 1993, Physica Scripta, T47, 133

\bibitem[{{Hu} {et~al.}(2011){Hu}, {Tout}, {Glebbeek}, \& {Dupret}}]{hu11}
{Hu}, H., {Tout}, C.~A., {Glebbeek}, E., \& {Dupret}, M.-A. 2011, \mnras, 418,
  195

\bibitem[{{Hui-Bon-Hoa}(2021)}]{huibonhoa21}
{Hui-Bon-Hoa}, A. 2021, arXiv e-prints, arXiv:2101.08510

\bibitem[{{Iglesias} \& {Rogers}(1996)}]{iglesias}
{Iglesias}, C.~A. \& {Rogers}, F.~J. 1996, \apj, 464, 943

\bibitem[{{LeBlanc} \& {Alecian}(2004)}]{leblanc_alecian}
{LeBlanc}, F. \& {Alecian}, G. 2004, \mnras, 352, 1329

\bibitem[{{Lebreton} {et~al.}(2008){Lebreton}, {Montalb{\'a}n},
  {Christensen-Dalsgaard}, {Roxburgh}, \& {Weiss}}]{lebreton08}
{Lebreton}, Y., {Montalb{\'a}n}, J., {Christensen-Dalsgaard}, J., {Roxburgh},
  I.~W., \& {Weiss}, A. 2008, \apss, 316, 187

\bibitem[{{Marques} {et~al.}(2013){Marques}, {Goupil}, {Lebreton}, {Talon},
  {Palacios}, {Belkacem}, {Ouazzani}, {Mosser}, {Moya}, {Morel}, {Pichon},
  {Mathis}, {Zahn}, {Turck-Chi{\`e}ze}, \& {Nghiem}}]{marques13}
{Marques}, J.~P., {Goupil}, M.~J., {Lebreton}, Y., {et~al.} 2013, \aap, 549,
  A74

\bibitem[{{Michaud} \& {Proffitt}(1993)}]{Michaud_f2}
{Michaud}, G. \& {Proffitt}, C.~R. 1993, in Astronomical Society of the Pacific
  Conference Series, Vol.~40, IAU Colloq. 137: Inside the Stars, ed. W.~W.
  {Weiss} \& A.~{Baglin}, 246--259

\bibitem[{{Michaud} {et~al.}(2011){Michaud}, {Richer}, \& {Vick}}]{michaud11}
{Michaud}, G., {Richer}, J., \& {Vick}, M. 2011, \aap, 534, A18

\bibitem[{{Morel} \& {Lebreton}(2008)}]{morel08}
{Morel}, P. \& {Lebreton}, Y. 2008, \apss, 316, 61

\bibitem[{{Morel} \& {Th{\'e}venin}(2002)}]{thevenin}
{Morel}, P. \& {Th{\'e}venin}, F. 2002, \aap, 390, 611

\bibitem[{{Nsamba} {et~al.}(2018){Nsamba}, {Campante}, {Monteiro}, {Cunha},
  {Rendle}, {Reese}, \& {Verma}}]{nsamba18}
{Nsamba}, B., {Campante}, T.~L., {Monteiro}, M.~J.~P.~F.~G., {et~al.} 2018,
  \mnras, 477, 5052

\bibitem[{{Paxton} {et~al.}(2011){Paxton}, {Bildsten}, {Dotter}, {Herwig},
  {Lesaffre}, \& {Timmes}}]{paxton1}
{Paxton}, B., {Bildsten}, L., {Dotter}, A., {et~al.} 2011, \apjs, 192, 3

\bibitem[{{Paxton} {et~al.}(2018){Paxton}, {Schwab}, {Bauer}, {Bildsten},
  {Blinnikov}, {Duffell}, {Farmer}, {Goldberg}, {Marchant}, {Sorokina},
  {Thoul}, {Townsend}, \& {Timmes}}]{paxton}
{Paxton}, B., {Schwab}, J., {Bauer}, E.~B., {et~al.} 2018, \apjs, 234, 34

\bibitem[{{Paxton} {et~al.}(2019){Paxton}, {Smolec}, {Schwab}, {Gautschy},
  {Bildsten}, {Cantiello}, {Dotter}, {Farmer}, {Goldberg}, {Jermyn}, {Kanbur},
  {Marchant}, {Thoul}, {Townsend}, {Wolf}, {Zhang}, \& {Timmes}}]{paxton2}
{Paxton}, B., {Smolec}, R., {Schwab}, J., {et~al.} 2019, \apjs, 243, 10

\bibitem[{{Rauer} {et~al.}(2014){Rauer}, {Catala}, {Aerts}, {Appourchaux},
  {Benz}, {Brandeker}, {Christensen-Dalsgaard}, {Deleuil}, {Gizon}, {Goupil},
  {G{\"u}del}, {Janot-Pacheco}, {Mas-Hesse}, {Pagano}, {Piotto}, {Pollacco},
  {Santos}, {Smith}, {Su{\'a}rez}, {Szab{\'o}}, {Udry}, {Adibekyan}, {Alibert},
  {Almenara}, {Amaro-Seoane}, {Eiff}, {Asplund}, {Antonello}, {Barnes},
  {Baudin}, {Belkacem}, {Bergemann}, {Bihain}, {Birch}, {Bonfils}, {Boisse},
  {Bonomo}, {Borsa}, {Brand {\~a}o}, {Brocato}, {Brun}, {Burleigh}, {Burston},
  {Cabrera}, {Cassisi}, {Chaplin}, {Charpinet}, {Chiappini}, {Church},
  {Csizmadia}, {Cunha}, {Damasso}, {Davies}, {Deeg}, {D{\'\i}az}, {Dreizler},
  {Dreyer}, {Eggenberger}, {Ehrenreich}, {Eigm{\"u}ller}, {Erikson}, {Farmer},
  {Feltzing}, {de Oliveira Fialho}, {Figueira}, {Forveille}, {Fridlund},
  {Garc{\'\i}a}, {Giommi}, {Giuffrida}, {Godolt}, {Gomes da Silva}, {Granzer},
  {Grenfell}, {Grotsch-Noels}, {G{\"u}nther}, {Haswell}, {Hatzes},
  {H{\'e}brard}, {Hekker}, {Helled}, {Heng}, {Jenkins}, {Johansen},
  {Khodachenko}, {Kislyakova}, {Kley}, {Kolb}, {Krivova}, {Kupka}, {Lammer},
  {Lanza}, {Lebreton}, {Magrin}, {Marcos-Arenal}, {Marrese}, {Marques},
  {Martins}, {Mathis}, {Mathur}, {Messina}, {Miglio}, {Montalban}, {Montalto},
  {Monteiro}, {Moradi}, {Moravveji}, {Mordasini}, {Morel}, {Mortier},
  {Nascimbeni}, {Nelson}, {Nielsen}, {Noack}, {Norton}, {Ofir}, {Oshagh},
  {Ouazzani}, {P{\'a}pics}, {Parro}, {Petit}, {Plez}, {Poretti}, {Quirrenbach},
  {Ragazzoni}, {Raimondo}, {Rainer}, {Reese}, {Redmer}, {Reffert},
  {Rojas-Ayala}, {Roxburgh}, {Salmon}, {Santerne}, {Schneider}, {Schou},
  {Schuh}, {Schunker}, {Silva-Valio}, {Silvotti}, {Skillen}, {Snellen}, {Sohl},
  {Sousa}, {Sozzetti}, {Stello}, {Strassmeier}, {{\v{S}}vanda}, {Szab{\'o}},
  {Tkachenko}, {Valencia}, {Van Grootel}, {Vauclair}, {Ventura}, {Wagner},
  {Walton}, {Weingrill}, {Werner}, {Wheatley}, \& {Zwintz}}]{rauer14}
{Rauer}, H., {Catala}, C., {Aerts}, C., {et~al.} 2014, Experimental Astronomy,
  38, 249

\bibitem[{{Richard} {et~al.}(2001){Richard}, {Michaud}, \&
  {Richer}}]{richard01}
{Richard}, O., {Michaud}, G., \& {Richer}, J. 2001, \apj, 558, 377

\bibitem[{{Richard} {et~al.}(2005){Richard}, {Michaud}, \&
  {Richer}}]{richard05}
{Richard}, O., {Michaud}, G., \& {Richer}, J. 2005, \apj, 619, 538

\bibitem[{{Richer} {et~al.}(1998){Richer}, {Michaud}, {Rogers}, {Iglesias},
  {Turcotte}, \& {LeBlanc}}]{richer98}
{Richer}, J., {Michaud}, G., {Rogers}, F., {et~al.} 1998, \apj, 492, 833

\bibitem[{{Richer} {et~al.}(2000){Richer}, {Michaud}, \& {Turcotte}}]{richer00}
{Richer}, J., {Michaud}, G., \& {Turcotte}, S. 2000, \apj, 529, 338

\bibitem[{{Ricker} {et~al.}(2015){Ricker}, {Winn}, {Vanderspek}, {Latham},
  {Bakos}, {Bean}, {Berta-Thompson}, {Brown}, {Buchhave}, {Butler}, {Butler},
  {Chaplin}, {Charbonneau}, {Christensen-Dalsgaard}, {Clampin}, {Deming},
  {Doty}, {De Lee}, {Dressing}, {Dunham}, {Endl}, {Fressin}, {Ge}, {Henning},
  {Holman}, {Howard}, {Ida}, {Jenkins}, {Jernigan}, {Johnson}, {Kaltenegger},
  {Kawai}, {Kjeldsen}, {Laughlin}, {Levine}, {Lin}, {Lissauer}, {MacQueen},
  {Marcy}, {McCullough}, {Morton}, {Narita}, {Paegert}, {Palle}, {Pepe},
  {Pepper}, {Quirrenbach}, {Rinehart}, {Sasselov}, {Sato}, {Seager},
  {Sozzetti}, {Stassun}, {Sullivan}, {Szentgyorgyi}, {Torres}, {Udry}, \&
  {Villasenor}}]{ricker15}
{Ricker}, G.~R., {Winn}, J.~N., {Vanderspek}, R., {et~al.} 2015, Journal of
  Astronomical Telescopes, Instruments, and Systems, 1, 014003

\bibitem[{{Rogers} \& {Nayfonov}(2002)}]{rogers02}
{Rogers}, F.~J. \& {Nayfonov}, A. 2002, \apj, 576, 1064

\bibitem[{{Seaton}(2005)}]{seaton052}
{Seaton}, M.~J. 2005, \mnras, 362, L1

\bibitem[{{Seaton} \& {Badnell}(2004)}]{seaton04}
{Seaton}, M.~J. \& {Badnell}, N.~R. 2004, \mnras, 354, 457

\bibitem[{{Semenova} {et~al.}(2020){Semenova}, {Bergemann}, {Deal},
  {Serenelli}, {Hansen}, {Gallagher}, {Bayo}, {Bensby}, {Bragaglia}, {Carraro},
  {Morbidelli}, {Pancino}, \& {Smiljanic}}]{semenova20}
{Semenova}, E., {Bergemann}, M., {Deal}, M., {et~al.} 2020, \aap, 643, A164

\bibitem[{{Th{\'e}ado} {et~al.}(2009){Th{\'e}ado}, {Vauclair}, {Alecian}, \&
  {LeBlanc}}]{theado09}
{Th{\'e}ado}, S., {Vauclair}, S., {Alecian}, G., \& {LeBlanc}, F. 2009, \apj,
  704, 1262

\bibitem[{{Thoul} {et~al.}(1994){Thoul}, {Bahcall}, \& {Loeb}}]{thoul}
{Thoul}, A.~A., {Bahcall}, J.~N., \& {Loeb}, A. 1994, \apj, 421, 828

\bibitem[{{Turcotte} {et~al.}(1998){Turcotte}, {Richer}, { }, {Iglesias}, \&
  {Rogers}}]{turcotte98}
{Turcotte}, S., {Richer}, J., { }, G., {Iglesias}, C.~A., \& {Rogers}, F.~J.
  1998, \apj, 504, 539

\bibitem[{{Verma} \& {Silva Aguirre}(2019)}]{verma19}
{Verma}, K. \& {Silva Aguirre}, V. 2019, \mnras, 489, 1850

\end{thebibliography}

\begin{appendix}
\section{Analysis of the impact of MESA atomic diffusion options on surface abundance profiles}
\label{Appendix_Opt}



In this appendix we address the impact of the options mentioned in Sect.~\ref{MESA_opsect} in more detail. We primarily focus on the physical significance of the code options, before presenting quantitative results applied to the models.
The list of these options, as well as their their default values, is presented in Table \ref{tab_tests}. The models are defined in Table \ref{mesa_models}, as well as a summary of the test results. Since some of these options have high computational costs, it may not be always feasible to choose values for these options that provide the maximum accuracy. Therefore, this appendix can be used as an indication, that should be adapted for each case, in order to reach the desired balance between the computational cost and the accuracy.


\subsection{{Surface mass and temperature cell control parameters}} \label{dq_dT}


This appendix is dedicated to
options 1 and 2 of Table \ref{tab_tests}, introduced in Sect. \ref{MESA_opsect}. In the limiting case where the option 1 is equal to the fraction of mass corresponding to the surface convective zone (model A11 in Fig.3), the whole zone will be treated as a single cell for the diffusion computation. Since convection completely homogenizes the chemical composition, the treatment of atomic diffusion is then done correctly. However, if this value is larger than the fraction of mass of the SCZ, errors may occur. In this case, the mixed region is artificially forced to be deeper, introducing an artificial and efficient full mixing in regions where it should not be present (see models A12 and A13 in Fig.3) and incorrectly uniformizing the composition of the star.



This option can be particularly critical in more massive stars, where the SCZ are very thin. This is why the default value is suitable as far as the size of the surface convective zone does not become smaller than this value. Analogously, one must not set a value for option 2 that is higher than the temperature at the bottom of the surface convective zone (see Fig.~\ref{dT}).

\begin{figure*}[b]
    \centering
   \includegraphics[scale=0.75]{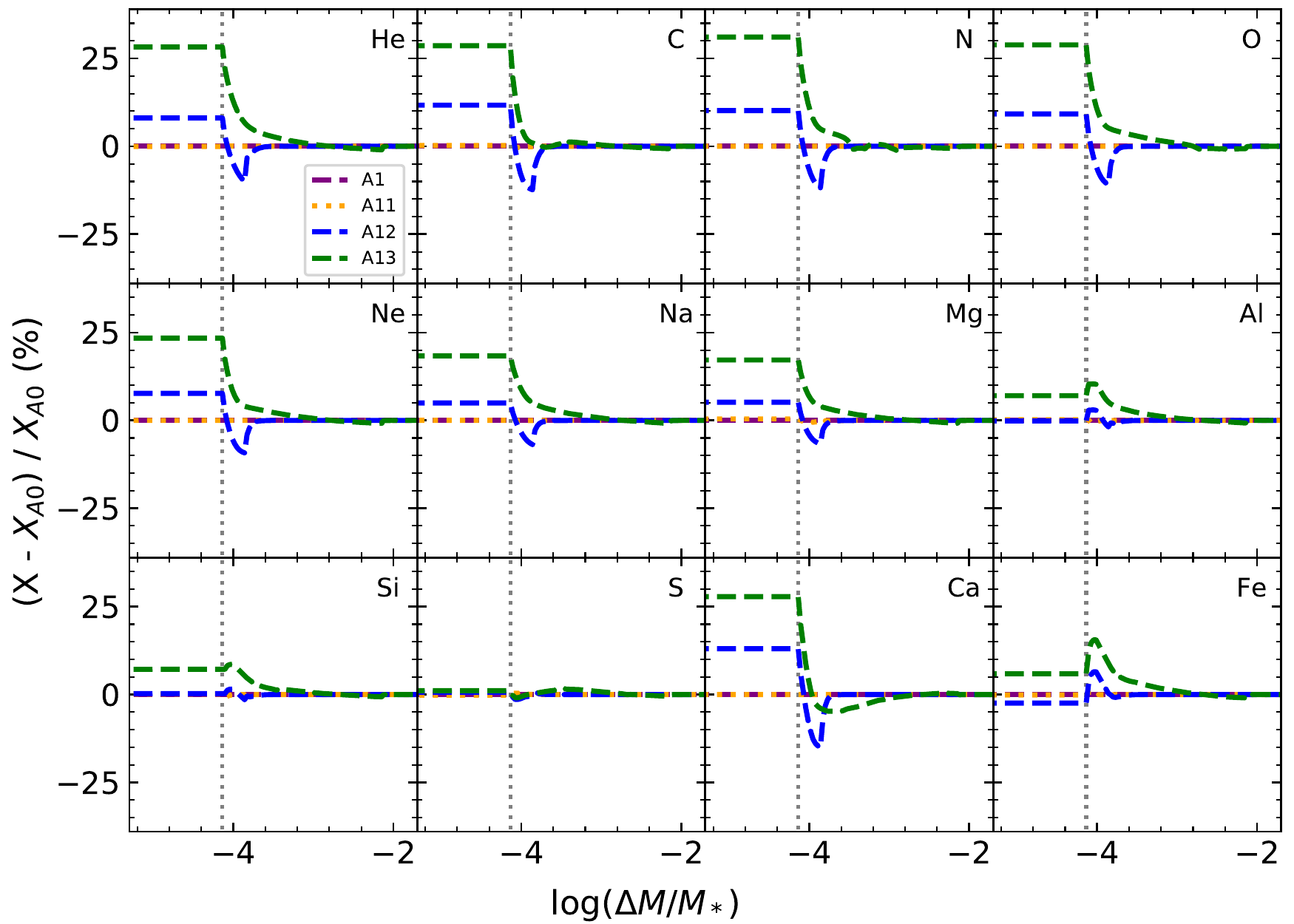} 
      \caption{Relative difference between the abundance profiles resulting from option 1 (\textit{diffusion\_min\_dq\_at\_surface}) tests at 420 Myr for 1.4$ M_\odot\ $ models and the A0 model. The option 1 values for each model are given in Table \ref{mesa_models}.
      The vertical dotted line represents the bottom of the surface convection zone of the A0 model.}
   \label{dq}
\end{figure*}
\begin{figure*}[t]
    \centering
   \includegraphics[scale=0.85]{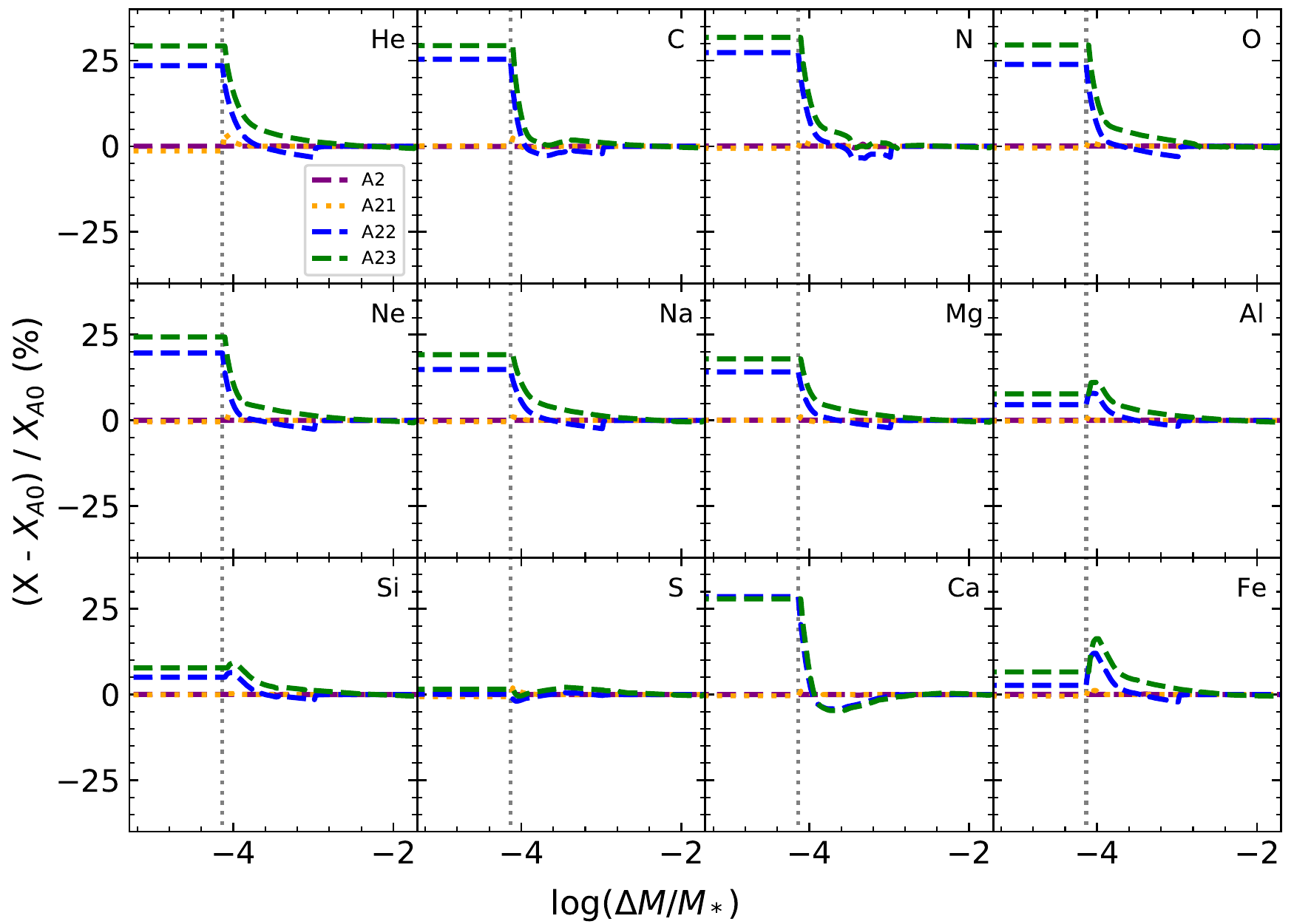} 
      \caption{Relative difference between the abundance profiles resulting from option 2 (\textit{diffusion\_min\_T\_at\_surface}) tests at 420 Myr for 1.4$ M_\odot\ $ models and the A0 model. The option 2 values for each model are given in Table \ref{mesa_models}. The vertical dotted line represents the bottom of the surface convection zone of the A0 model.}
   \label{dT}
\end{figure*}

Indeed, the abundance differences caused by this artificial full mixing are already significant for the A12 and A22 models (Figs. \ref{dq} and \ref{dT}), where the values of the options 1 and 2 are twice the mass of the SCZ and the temperature at its bottom, respectively.


The average surface abundance differences in terms of mass fraction between these models and A0 are about 6\% for A12, and 16\% for A22 (Table~\ref{mesa_models}). 
For $^{12}$C, $^{14}$N and $^{40}$Ca, these differences are almost doubled. This translates to a surface abundance difference up to 0.06~dex for the A12 model, 0.14~dex for the A22 model and 0.16~dex for models A13 and A23 (see Fig.~\ref{fig:diff-test}).


It should also be noticed that setting options 1 or 2 to $M_{SCZ}$ or $T_{BSCZ}$, respectively, allows to save up to 15\% of the computation time, compared to the default settings, without affecting significantly the abundance profiles. Moreover, the deeper the SCZ is, the larger the value of options 1 and 2 can be, hence a faster computation. This is particularly interesting for the less massive models.


\subsection{Opacity upper temperature limit}
\label{opacity}

The computation of the Rosseland mean opacity using OPCD3 monochromatic tables, in the A0 model, was restricted to the temperature interval $[10^4, 10^{6.3}]$~K. Outside this interval, opacity tables with a fixed metal mixture are used. This range is defined by options 3 to 6 in the following manner: if log(T) is in the interval [$O_6,O_4$], where  $O_i$ represents the value of the \textit{ith} option, "OP\_mono" opacities will be used. Standard opacity tables are used if $log(T) > O_3$ or $log(T) < O_5$, and in the remaining intervals both are partially used. 
In the models of this paper, given that $O_3 = O_4$ and $O_5 = O_6$, the "OP\_mono" opacities are either fully used or not at all. The tests carried out in this appendix involved changing the upper limit (options 3 and 4, simultaneously).

This possibility of adapting the temperature range where monochromatic opacities are used can be a way to save computational time. In the regions of the model where the mixture of metal is not strongly affected by atomic diffusion, it is a reasonable approximation (see \citealt{huibonhoa21} for more massive stars). In the following, we quantify its effect on the surface abundances.

\begin{table*}[b]
\setlength{\tabcolsep}{5pt}
\caption{\textsc{MESA} options addressed in this paper.}
\label{tab_tests}
\begin{adjustbox}{width=18.5cm,center}
\centering
\begin{tabular}{l|l|l|l|l}
\hline\hline
& Option name   & Brief explanation & Default value & A0 value   \\
\hline 
1 & diffusion\_min\_dq\_at\_surface  &  At least that much fraction of star 
mass gets treated &  $10^{-9}$ & $5 \times 10^{-9}$\\
& & as a single cell. & & \\
\hline
2 & diffusion\_min\_T\_at\_surface     & At least that much fraction of star with $T < $ value gets   & $10^4$ & $5 \times 10^{4}$ \\
&   & treated as a single cell. & &\\

\hline 
3 & high\_logT\_op\_mono\_full\_off  & Outside of the temperature range defined by these & $-10^{99}$   & 6.3  \\  
4 & high\_logT\_op\_mono\_full\_on & options, the code will use standard opacity tables    & $-10^{99}$  & 6.3 \\
5 & low\_logT\_op\_mono\_full\_off & instead of "OP\_mono" ones (see Sect. \ref{opacity}). & $-10^{99}$ & 4.0 \\
6 & low\_logT\_op\_mono\_full\_on & Value order: $3\geqslant4\geqslant5\geqslant6$ & $-10^{99}$ & 4.0 \\
\hline
7 & radiation\_turbulence\_coeff & Introduces turbulent mixing. & 0 & 0   \\
 \hline 
8 & diffusion\_num\_classes & Number of representative classes of species for diffusion    & 5 & 14 \\
&  & calculations. & & \\
 \hline
 9 & diffusion\_class\_representative(:) &  Isotope names for diffusion representatives.  & $^1$H, $^3$He, $^4$He, $^{16}$O,$^{56}$Fe  & Sect. \ref{MESA}\\
 \hline
 10 & diffusion\_class\_A\_max(:)  & Defines the diffusion classes (A is atomic mass). The   & 2, 3, 4, 16, $10^4$ & Sect. \ref{ab_compar}\\
  &   & species goes into the 1st class with $A\_max \leq$ species A. & & \\
\hline 
\end{tabular}%
\end{adjustbox}
\label{options}
\end{table*}

\begin{table*}
\setlength{\tabcolsep}{5pt}
\caption{\textsc{MESA} models description, using the options defined in Table \ref{tab_tests}. Both $M_{SCZ}$ and $T_{BSCZ}$ are defined using the highest position of the SCZ in the stars, in order to avoid the issue described in Sect.~\ref{dq_dT}. The value for option 3 of the model A3 (default value) makes it so that only standard opacity tables are used.}
	\label{mesa_models}
\centering
\begin{tabular}{l|l|l|l|l}
\hline\hline
Model & Option & Option value & Surface ab. diff. \tablefoottext{c} (\%) & Max. surface ab. diff. \tablefoottext{d} (dex)  \\[2pt]
\hline 
A0 &   &  $5 \times 10^{-9} M_\star$ & --- & ---\\  
A1  &  & $1 \times 10^{-9} M_\star$ (default) &  0.00  & 0.00  \\  
A11 &  1 &  $M_{SCZ}$ ($7.30 \times 10^{-5} M_\star$)  & 0.05  & 0.00  \\  
A12 &  & $2M_{SCZ}$   &  5.69  & 0.06 \\  
A13 & & $100M_{SCZ}$   & 18.74  & 0.16  \\   
\hline
A0 & &  $5 \times 10^{4} K$    & ---  & ---  \\ 
A2 & & $1 \times 10^{4} K$ (default)  &  0.00  & 0.00 \\  
A21 & 2  & $T_{BSCZ}$ ($5.01  \times 10^{5} K$)  & 0.39  & 0.00 \\ 
A22 & & $2T_{BSCZ}$   & 15.82  & 0.14  \\ 
A23 & & $5T_{BSCZ}$   &  19.42  &  0.16 \\ 
\hline 
A0 &   & 6.3  & ---  & --- \\  
A3 &  3 and 4 & $-1 \times 10^{99}$ (default) \tablefoottext{a} & 0.94  &  0.01 \\ 
A31  &  & 5.8  & 0.85   & 0.01 \\ 
A32 &  & 7.3\tablefoottext{b}   & 3.08  & 0.03 \\ 
\hline 
\end{tabular}%
\label{mod_set}
\tablefoot{\tablefoottext{a}{Due to the value order presented in Table 2, the options 5 and 6 were also set to their default values.}
\tablefoottext{b}{As $\log_{10}(T_\mathrm{core})=7.25$, the A32 model takes monochromatic opacities into account in the whole interior in this case}.\tablefoottext{c}{Mean surface abundance difference in mass fraction over all elements, relative to A0.} \tablefoottext{d}{Maximum surface abundance difference over all elements, relative to A0} (see Fig. \ref{fig:diff-test}).})
\end{table*}

\begin{figure*}[ht][b]
    \centering
   \includegraphics[scale=0.85]{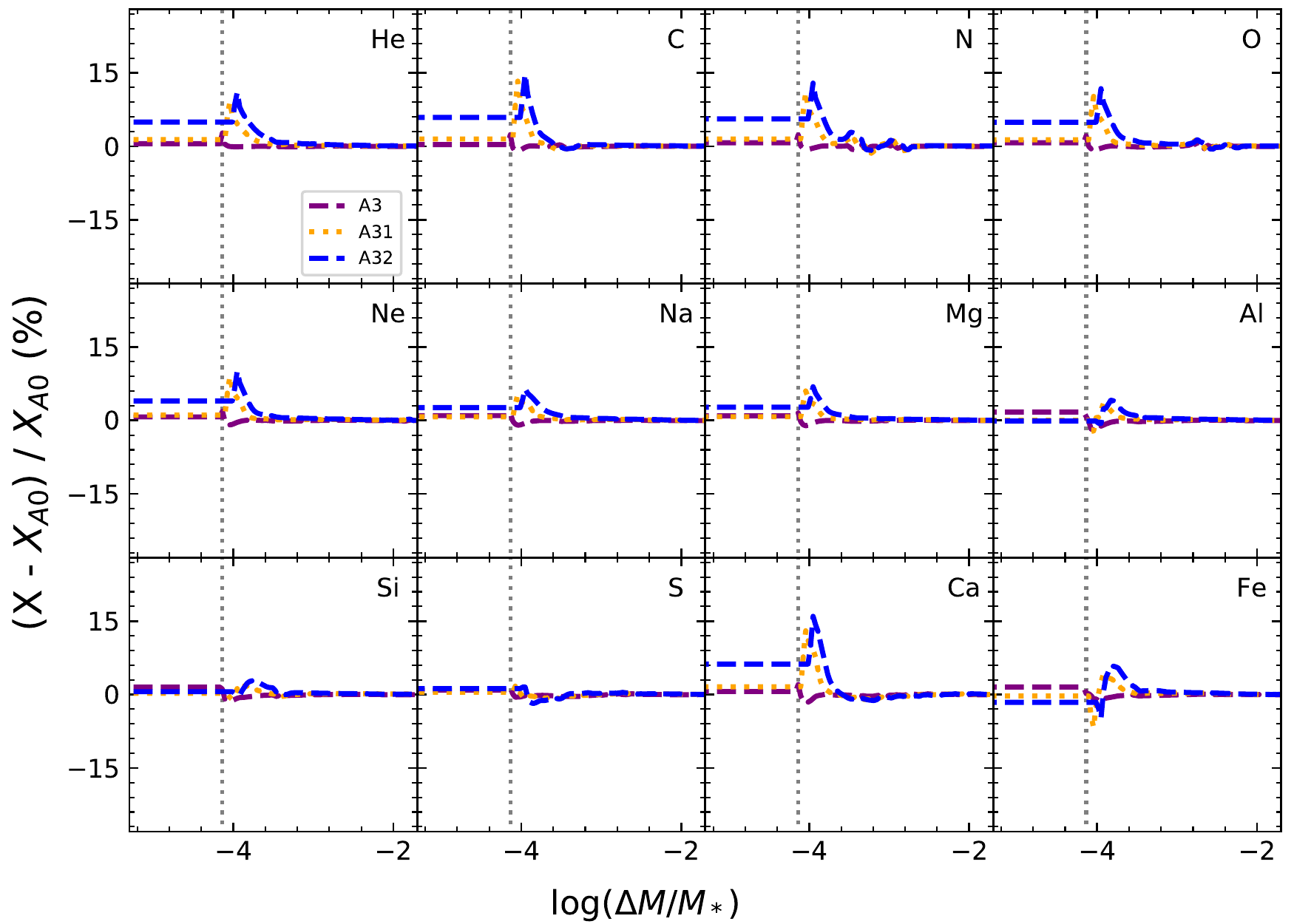} 
      \caption{Relative difference between the abundance profiles resulting from varying option 3 (\textit{high\_logT\_op\_mono\_full\_off}) at 410 Myr for 1.4$ M_\odot\ $ models and the A0 model. The models are defined in Table \ref{mesa_models}. The vertical line represents the bottom of the SCZ of the A0 model.}
   \label{high_logt}
\end{figure*}

Figure \ref{high_logt} shows the relative differences between the abundance profiles of models including larger temperature intervals for the use of monochromatic opacities and the A0 model. The main impact of altering the temperature range is to slightly modify the stellar structure in the regions where the opacity tables are different (T > $10^{6.3}$~K). This has a limited effect on the surface abundances, which indicates that for a reasonable accuracy/computation time compromise, an upper temperature of T=$10^{6.3}$~K is adequate. It is important to note that changing this upper temperature value has an impact on the mixing length parameter obtained with a solar calibration. Hence, these differences in surface abundance may be even smaller.

Indeed, the average surface abundance difference between the A0 and A3 models is only 0.9\%, while it is 4.1\% between the A0 and A32. This translates into 0.008 and 0.04~dex, respectively (See Fig.~\ref{fig:diff-test}). Extending the upper limit to $10^{7.3}$~K doubles the A0 model computation time. Furthermore, the A3 model, which only uses standard opacity tables, is about four times faster than the A0 one.

Lastly, while we have discussed the effects of altering the upper limit for opacity calculations using monochromatic opacities (options 3 and 4), there is also a lower temperature limit ($10^4$K, options 5 and 6) implemented in the A0 model. We decided not to change it since, at lower temperatures, molecules would start to have a significant impact on the opacity. Given that their contribution is not taken into account in the current version of OP monochromatic opacities, it is more reasonable to use standard opacity tables in this temperature range.


\subsection{Extra options}
In this appendix we discuss the effects of two additional options that, while not directly related to atomic diffusion, are nevertheless worth mentioning, given that they can significantly alter the surface abundances of the models.

\subsubsection{Radiation turbulence coefficient}
\label{turb_chapter}


Turbulent mixing counteracts the depletion of helium and metals in the outer envelope of stars. Radiative viscosity was thought to induce turbulent mixing of chemical elements \citep{thevenin}. However, the description of the effect of radiative viscosity on chemical element transport, as it was described in that paper, has been shown to be erroneous \citep{alecian}.

It is possible to include this effect with an option available in \textsc{MESA}, i.e. option 7 (\textit{radiative\_turbulence\_coeff}) of Table~\ref{tab_tests}. In order to test its effects we adopted two values: 0 (the default one) and 1. The average impact on the surface abundances is 4.2\% (up to $9$\% for $^{12}$C and $^{40}$Ca). 
Nevertheless, since this process is no longer physically justified \citep{alecian}, its usage is not recommended.
\subsubsection{Diffusion classes}
\label{diff_class}
In \textsc{MESA}, the treatment atomic diffusion is done with element classes (options 8, 9 and 10 of Table~\ref{tab_tests}), which give the possibility to group elements and follow their evolution as a single one. They define the number of representative classes of species for diffusion calculations, as well as their atomic mass. In the A0 models, all considered elements were associated with their own class. This is the only way to have atomic diffusion treated properly for these elements. These options are convenient when some elements are not needed for specific analysis, because they save computational time. In order to avoid any inconsistency in the opacity profiles of the models, the usage of one class per element is recommended.

\FloatBarrier
\section{Comparison between MESA, the Montreal/Montpellier code and CESTAM modes at 1.1~M$_\odot$}

\begin{figure}[h!]
    \centering
    \includegraphics[scale=0.65]{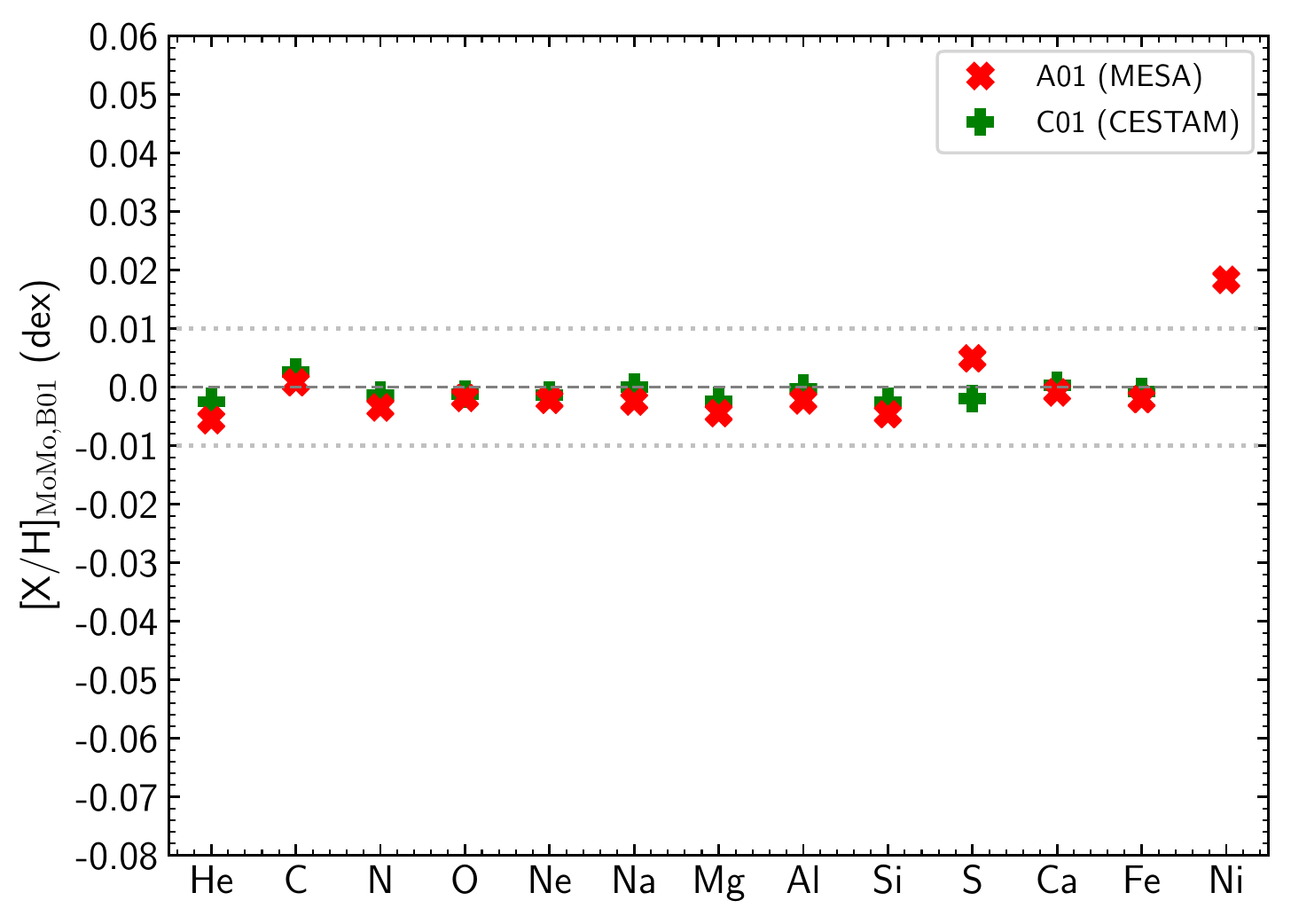}
     \caption{Same legend as Fig.~\ref{1.4-reldiff} for models A01 and C01.}
  \label{compar-1.1v2}
\end{figure}

\begin{figure*}[h!]
  \centering
  \includegraphics[scale=0.75]{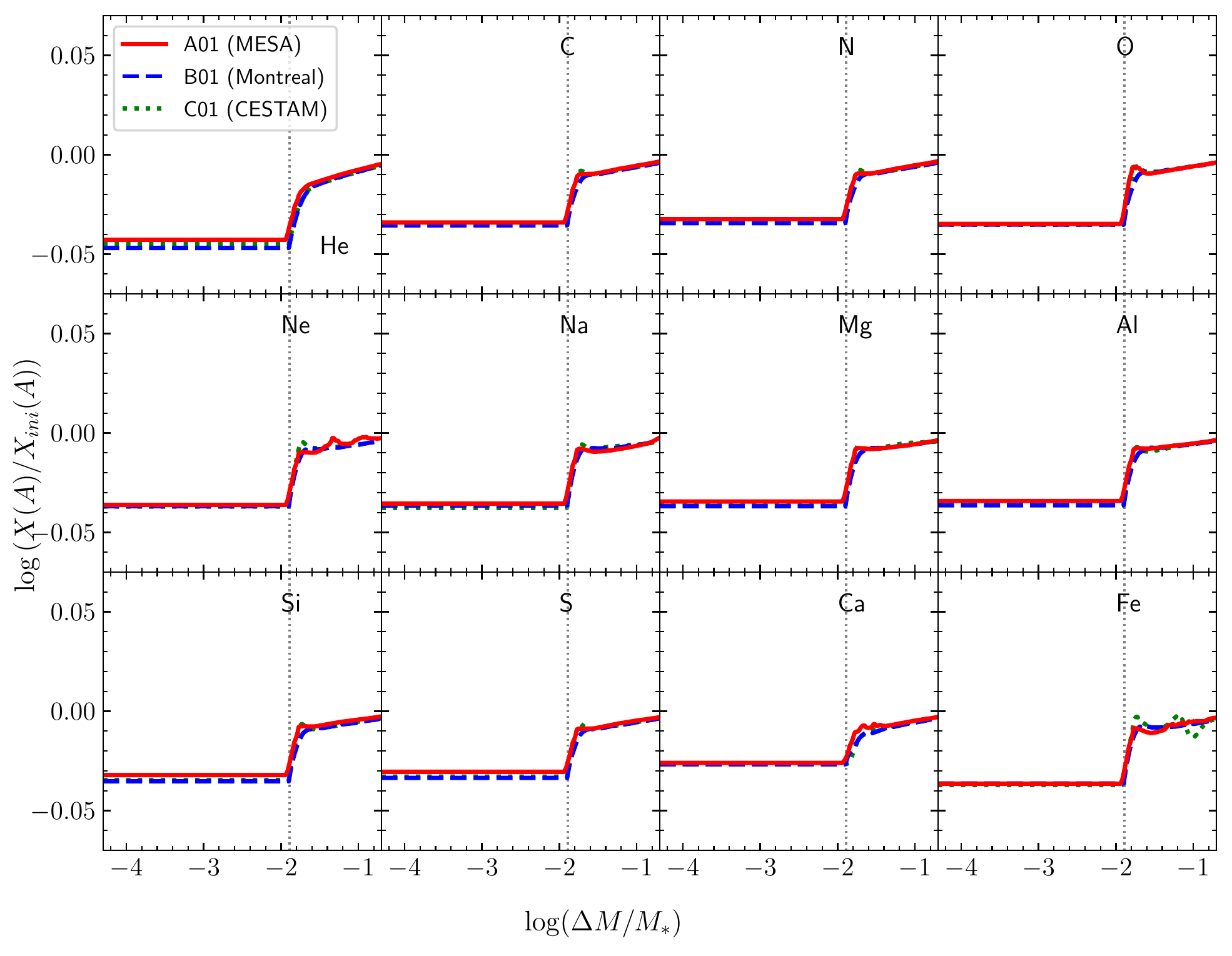}
  \caption{Same legend as Fig.~\ref{compar} for models A01, B01 and C01.}
  \label{compar-1.1}
\end{figure*}

This appendix is dedicated to the comparison between the three codes for models A01, B01 and C01. The comparison of the abundance profiles is presented in Fig.~\ref{compar-1.1}. The relative difference between the Montreal/Montpellier model and the two other codes is presented in Fig.~\ref{compar-1.1v2}.

\nopagebreak
\FloatBarrier

\section{Radiative acceleration profiles for A0, B0 and C0 models}

\begin{figure}[H]
\center\onecolumn\includegraphics[scale=0.75]{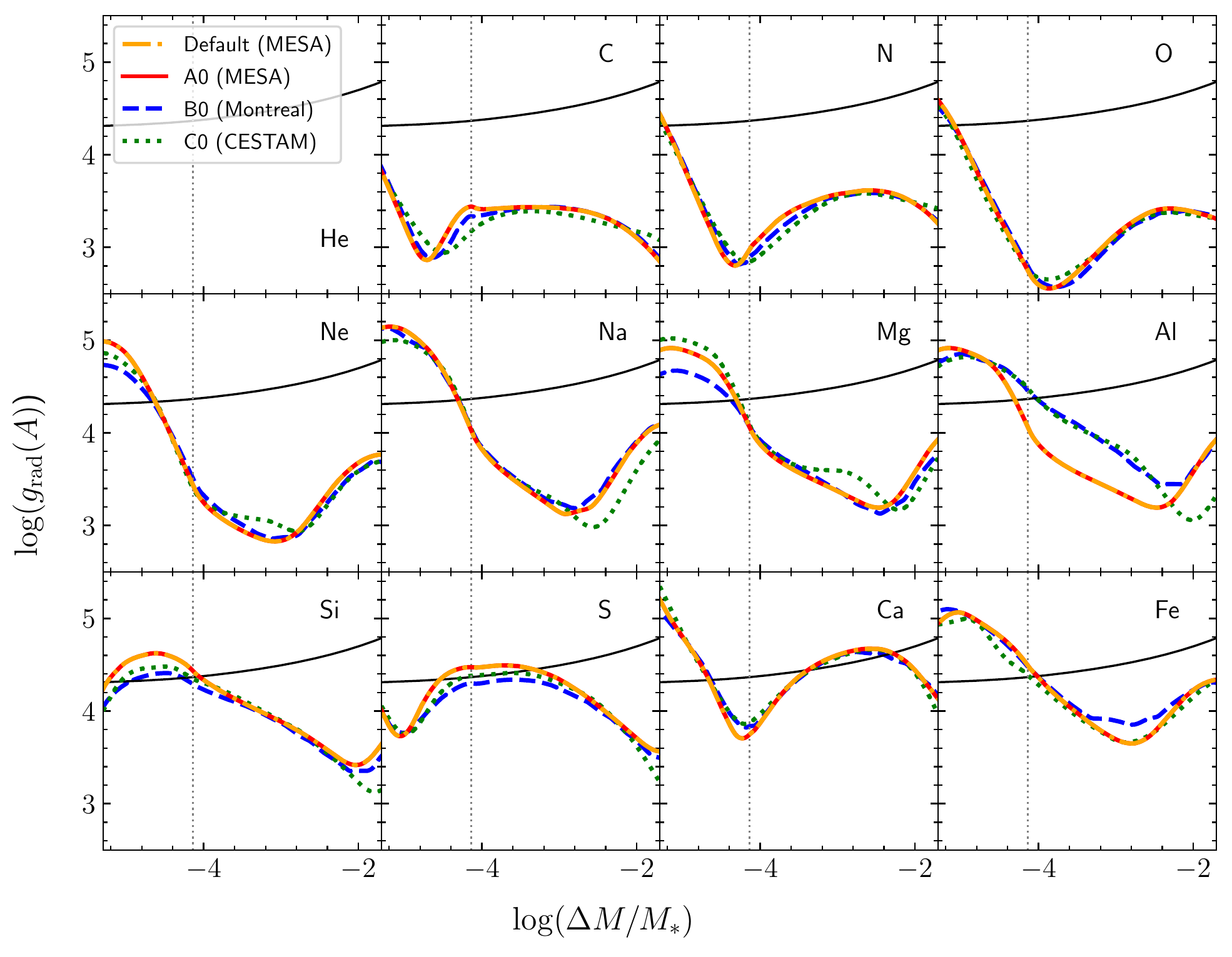}
      \caption{Radiative acceleration profiles according to $\log(\Delta M/M_\ast)$ (with $\Delta M$ being the mass above the radius r) obtained with \textsc{MESA} (A0 model, red solid lines), the Montreal/Montpellier code (B0 model, blue dashed lines) and \textsc{CESTAM} (C0 model, green dotted lines). The black solid line represent the gravity. The vertical grey dotted line represents the average position of the surface convective zone of the three models. The surface is towards the left part of the plots.}
   \label{compar-grad}
\end{figure}

In this appendix we present a comparison of the radiative acceleration profiles for models A0, B0 and C0 (Fig. \ref{compar-grad}).

\twocolumn

\FloatBarrier
\clearpage
\section{Surface abundance differences between A0 and, A1X, A2X and A3X models}

\begin{figure*}[b]
\label{fig:final}
    \centering
    \includegraphics[height = 190pt]{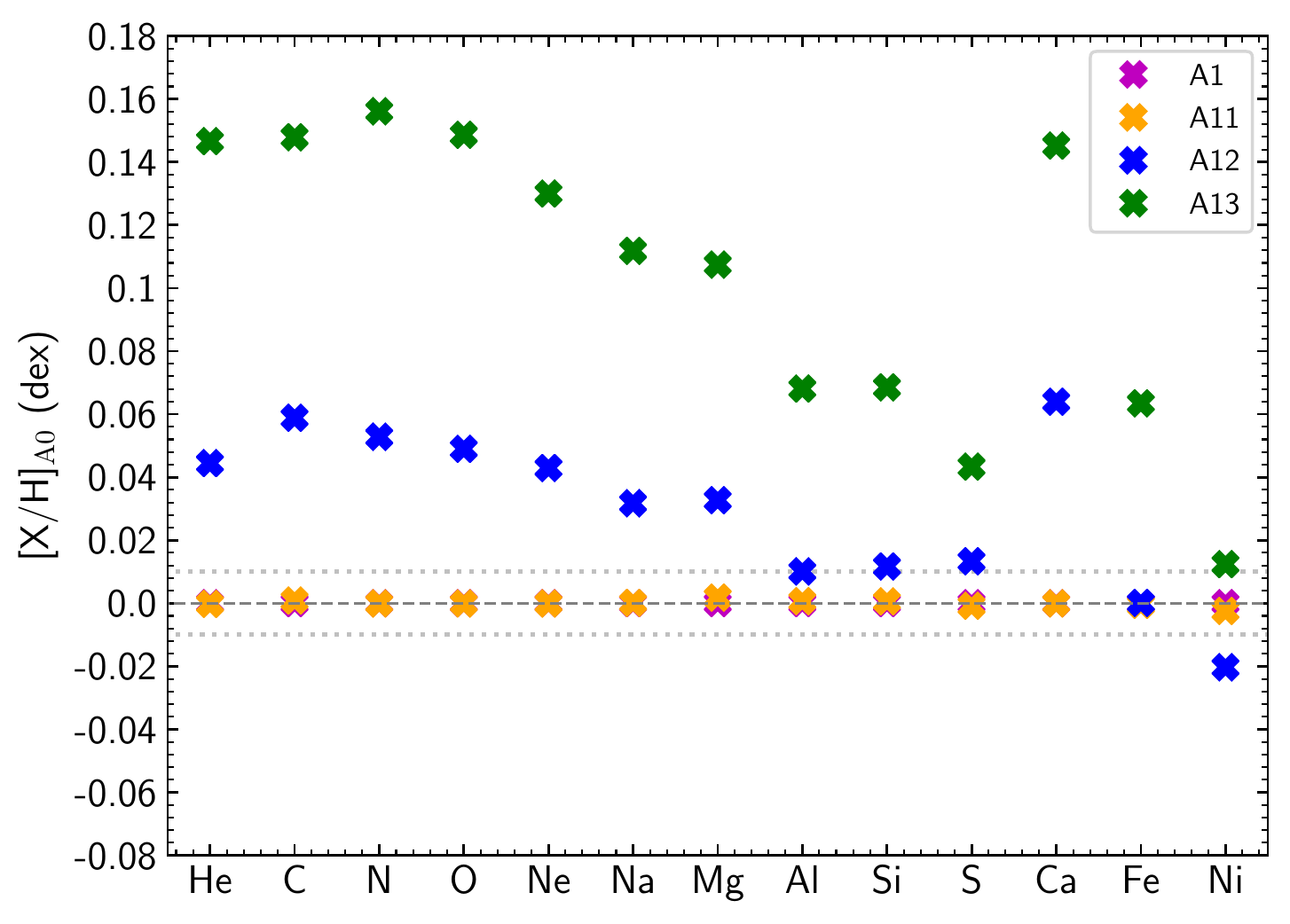}
    \includegraphics[height = 190pt]{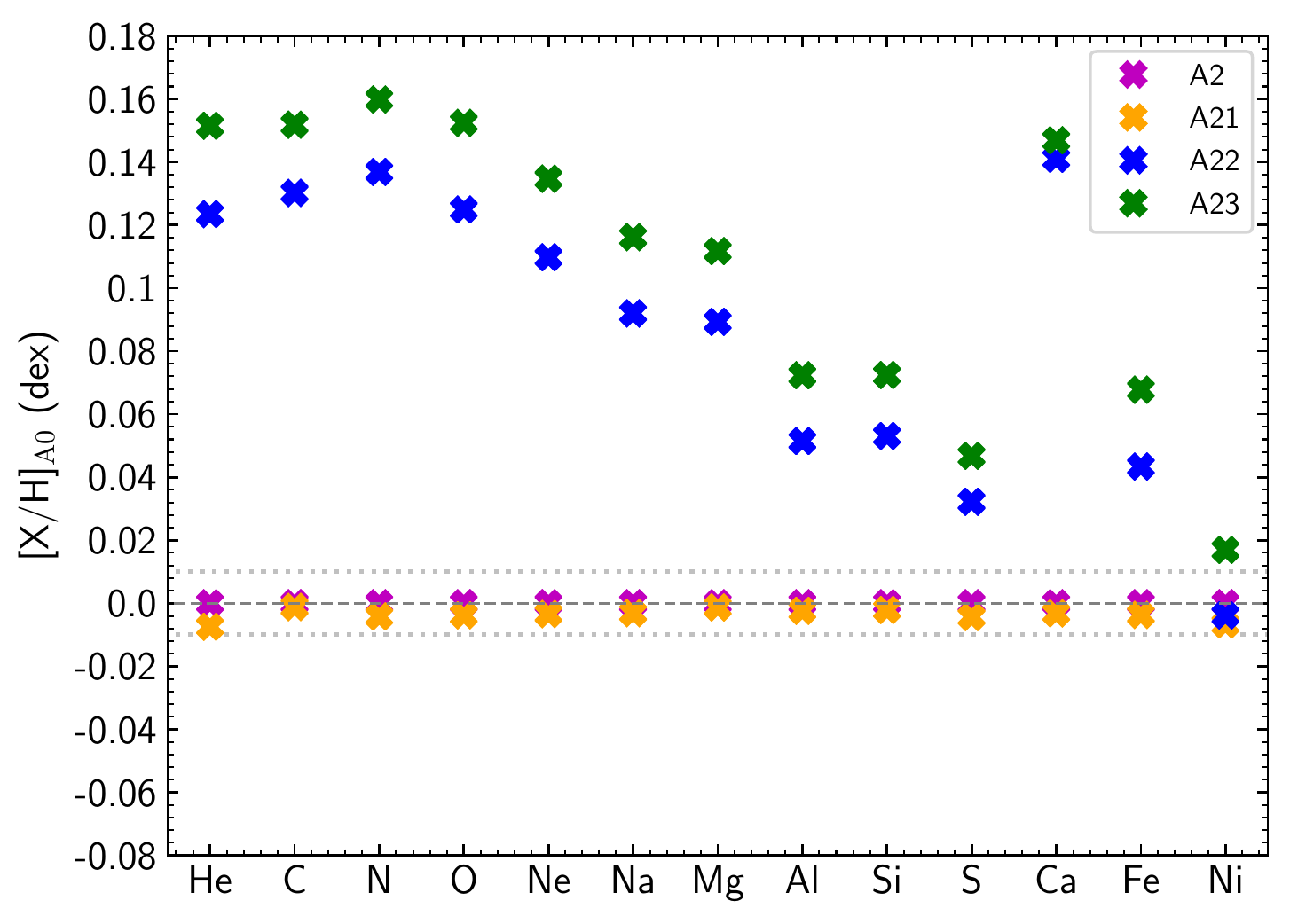}
    \includegraphics[height = 190pt]{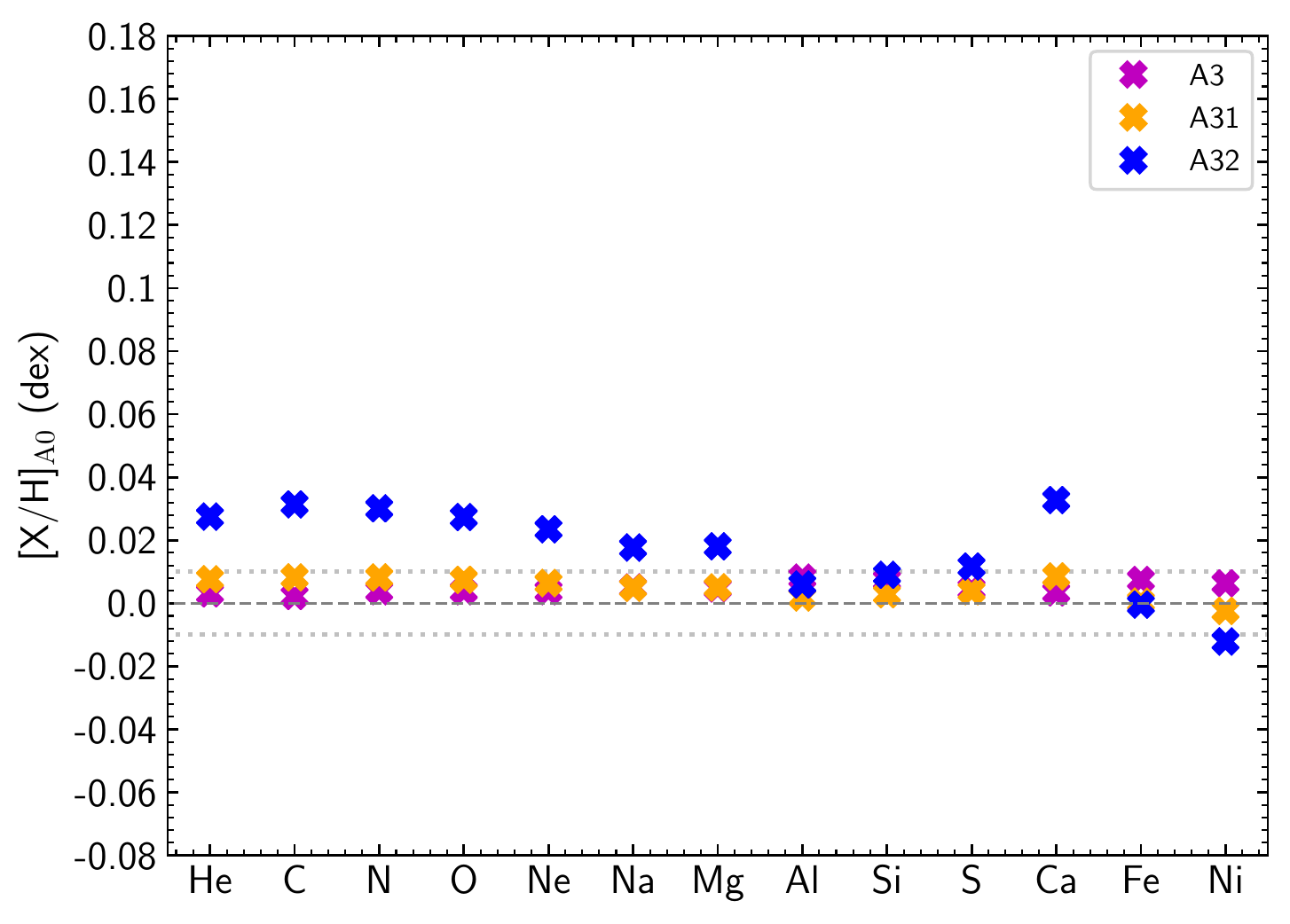}
      \caption{Difference of surface abundances between the model A0 and all the models presented in Table~\ref{tab_tests}. The grey dotted lines show the $\pm0.01$~dex interval.}
   \label{fig:diff-test}
\end{figure*}

In this appendix we present the surface abundances differences (in dex) between A0 and all models in Table~\ref{tab_tests}. In Figure \ref{fig:final}, [A/H]$_\mathrm{A0}=\log_{10}\left(\frac{X(A)}{X(H)}\right)_\mathrm{AX}-\log_{10}\left(\frac{X(A)}{X(H)}\right)_\mathrm{A0}$, where $X(A)$ and X(H) are the mass fractions of the element A and hydrogen, respectively.


\end{appendix}

\end{document}